\documentclass[aps,prd,superscriptaddress,notitlepage,showpacs]{revtex4-1}

\usepackage{graphicx}
\usepackage{amsmath}
\usepackage{amssymb}
\usepackage{color}
\usepackage{bm}

\newcommand{\sign}{\,\mbox{sign}}

\begin{document}

\title{Quantum oscillations as a probe of interaction effects in Weyl semimetals in a magnetic field}
\date{September 17, 2014}

\author{E. V. Gorbar}
\affiliation{Department of Physics, Taras Shevchenko National Kiev University, Kiev, 03680, Ukraine}
\affiliation{Bogolyubov Institute for Theoretical Physics, Kiev, 03680, Ukraine}

\author{V. A. Miransky}
\affiliation{Department of Applied Mathematics, Western University, London, Ontario N6A 5B7, Canada}

\author{I. A. Shovkovy}
\affiliation{College of Letters and Sciences, Arizona State University, Mesa, Arizona 85212, USA}

\author{P. O. Sukhachov}
\affiliation{Department of Physics, Taras Shevchenko National Kiev University, Kiev, 03680, Ukraine}

\begin{abstract}
The Weyl semimetal surface is modeled by applying the Bogolyubov boundary conditions, 
in which the quasiparticles have an infinite Dirac mass outside the semimetal. For a Weyl 
semimetal shaped as a slab of finite thickness, we derive an exact spectral equation for the 
quasiparticle states and obtain the spectrum of the bulk as well as surface Fermi arc modes. 
We also show that, in the presence of the magnetic field, the separation between Weyl 
nodes in momentum space and the length of the Fermi arcs in the reciprocal space are 
affected by the interactions. As a result, we find that the period of oscillations of the 
density of states related to closed magnetic orbits involving Fermi arcs has a nontrivial 
dependence on the orientation of the magnetic field projection in the plane of the semimetal 
surface. We conclude that the momentum space separation between Weyl nodes and its 
modification due the interaction effects in the magnetic field can be measured in the 
experimental studies of quantum oscillations.
\end{abstract}


\maketitle

\section{Introduction}
\label{Intro}

It was recently suggested that Weyl semimetals whose low energy electron excitations are 
described by the Weyl equation may be realized in the pyrochlore iridates \cite{Savrasov}. 
Another material HgCr$_2$Se$_4$ was later proposed as well \cite{Weng}. 
A Weyl semimetal phase may be realized also in topological insulator
heterostructures \cite{Balents} and in magnetically doped topological insulators
\cite{Cho}. Although not experimentally observed yet, Weyl semimetals have been
very actively studied theoretically (for reviews, see Refs.~\cite{Hook,Turner,Vafek}).
The recent experimental observation of Dirac semimetals \cite{Liu,Neupane,Borisenko}
raises the prospects for the observation of Weyl semimetals as well.

Because of the chiral symmetry in the Weyl equation, Weyl points
cannot be eliminated by perturbations as long as the translation symmetry is preserved,
realizing a topological phase of matter \cite{Volovik}. Indeed, nontrivial topological properties
of the Weyl nodes are related to the fact that they are monopoles of a Berry flux in the
Brillouin zone.

It was shown already a long time ago by Nielsen and Ninomiya
\cite{Nielsen} that Weyl nodes in crystals occur only in pairs of opposite chirality.
Therefore, the simplest way to eliminate a pair of Weyl nodes is to decrease the
separation between them in the momentum space transforming the Weyl nodes
into a Dirac point when they meet.

Negative longitudinal magnetoresistivity \cite{NN,Aji,Son} of Weyl semimetals is a
fascinating consequence of the chiral anomaly \cite{anomaly} (note that the longitudinal
magnetoresistivity is negative at sufficiently large magnetic fields for Dirac semimetals
too \cite{magnetoresistivity}). There are also other ways of probing the chiral anomaly
in Weyl semimetals connected with nonlocal transport \cite{Parameswaran}, coupling
between collective modes \cite{Qi}, nonanalytic nonclassical correction to the electron
compressibility and the plasmon frequency \cite{Panfilov}, the nonvanishing gyrotropy
induced by external fields \cite{Hosur-Qi,Tewari}, and optical absorption \cite{Ashby-Carbotte}.
Electromagnetic response of Weyl semimetals was considered in Refs.~\cite{Ran,Turner,Vafek,Hosur}.
We would like to mention also that one should be careful in drawing conclusions
on the consequences of the chiral anomaly in Weyl semimetals (see a discussion
of the chiral magnetic effect in Refs.~\cite{Franz,Basar,Chen}).

The topological nature of Weyl nodes leads to the surface Fermi arc states 
\cite{Savrasov,Haldane,Okugawa} that connect Weyl nodes of opposite chirality. These surface states 
are topologically protected and are well defined at momenta away from the Weyl nodes
because there are no bulk states with the same energy and momenta. Taken together
the two Fermi arc states on opposite surfaces form a closed Fermi surface.

The electron-electron interactions are important in Weyl semimetals and may lead to
the dynamical chiral symmetry breaking with the gap generation due to the pairing of
electrons and holes from Weyl nodes of opposite chirality \cite{Ran,Wei,Wang,Sukhachov}.
Recently, using the framework of a relativistic Nambu--Jona-Lasinio model
\cite{NJL-model1,NJL-model2}, three of us have shown \cite{engineering} that interaction
effects lead to dramatic consequences in a Dirac semimetal at nonzero charge
density in a magnetic field, making it a Weyl semimetal with a pair of Weyl nodes
for each of the original Dirac points. The Weyl nodes are separated in momentum
space by a dynamically generated chiral shift $\mathbf{b}$ directed along the magnetic field.
The magnitude of the momentum space separation between the Weyl nodes is
determined by the quasiparticle charge density, the strength of the magnetic field,
and the strength of the interaction. Although simple models with contact four-fermion
interaction were considered in Refs.~\cite{NJL-model1,NJL-model2,engineering}, the study performed
in QED \cite{QED} indicates the universality of the dynamical chiral shift generation
in a magnetic field and at a nonzero charge density. Experimentally, the transition from a
Dirac semimetal to a Weyl one in a magnetic field might have been observed in Bi$_{1-x}$Sb$_x$ for
$x \approx 0.03$ \cite{Kitaura}. Using the Hubbard model, it was shown in Ref.~\cite{Abanin}
that the interaction effects change the momentum space separation between the
Weyl nodes also in the absence of the magnetic field.

It should be emphasized that, in the context of our study, one should distinguish the 
following two types of interaction effects: (i) those that exist even in the absence 
of the magnetic field, (ii) those that are triggered by the field. While the chiral shift 
is already renormalized by interaction in the absence of the magnetic field \cite{Abanin},
we will concentrate exclusively on the effects intimately connected with the presence 
of both the magnetic field and the interaction.

Clearly, the chiral shift, separating the Weyl nodes in the momentum space, is a parameter which defines
a Weyl semimetal. Therefore, it is very desirable to have experimental means to determine it.
An anomalous nonquantized quantum Hall effect \cite{Weng,Balents,magnetoresistivity,Zyuzin,Grushin,Goswami}
presents one such means. A remarkable new way to determine experimentally the
momentum space separation of Weyl nodes was suggested in Ref.~\cite{Vishwanath}.
Although the surface states of Weyl semimetals consist of disjoint Fermi arcs,
it was shown that there exist closed magnetic orbits involving
the surface Fermi arcs. These orbits produce periodic quantum oscillations of the density
of states in a magnetic field. If observed, this unconventional Fermiology of surface states
would provide a clear fingerprint of the Weyl semimetal phase. Since, according to
Ref.~\cite{engineering}, the interaction effects change the separation of Weyl nodes
in momentum space, they may affect the quantum oscillations in Weyl semimetals.
This problem provides the main motivation for the present study.

This paper is organized as follows. The Weyl semimetal model is introduced in
Sec.~\ref{Model}. By utilizing the Bogolyubov model, we determine the surface 
Fermi arcs in Sec.~\ref{FermiArcs}. The effect of interaction on the chiral shift 
of the Weyl semimetal in a magnetic field is studied in Sec.~\ref{Interaction}. 
The analysis of the quantum oscillations associated with the surface Fermi 
arcs is performed in Sec.~\ref{oscillations}, where we reproduce the original 
derivation of the quantum $1/B$ oscillations given in Ref.~\cite{Vishwanath} 
and then extend it to the case of an interacting system in a magnetic field. 
In Sec.~\ref{SurfaceBulk} a superposition of bulk and Fermi-arc related 
types of quantum oscillations in bulk probes is briefly investigated. The discussion 
of the main results is given in Sec.~\ref{Discussion}. Some technical details 
of the analysis are presented in appendices at the end of the paper.

Throughout this paper, we use the units with $\hbar=1$ and $c=1$.

\section{Model of Weyl semimetal with infinite mass boundary condition}
\label{Model}

The low-energy Hamiltonian of a Weyl semimetal with two Weyl nodes of opposite
chiralities in the bulk has the following form:
\begin{equation}
H=H_0+H_{\rm int},
\label{Hamiltonian-model-Weyl}
\end{equation}
where
\begin{equation}
H_0=\int d^3r \,v_F \psi^{\dagger} (\mathbf{r})\left(
\begin{array}{cc} \bm{\sigma}\cdot(-i\bm{\nabla}-\mathbf{b}_{0}) -\mu_0 & 0\\ 0 &
-\bm{\sigma}\cdot(-i\bm{\nabla}+\mathbf{b}_{0}) -\mu_0 \end{array}
\right)\psi(\mathbf{r})
\label{free-Hamiltonian}
\end{equation}
is the free Hamiltonian which describes two Weyl nodes of opposite chiralities, separated by
vector $2\mathbf{b}_{0}$ in momentum space, $v_F$ is the Fermi velocity, 
$\mu_0$ is the chemical potential, and $\bm{\sigma}=(\sigma_x,\sigma_y,\sigma_z)$ 
are Pauli matrices associated with
the band degrees of freedom \cite{Burkov3,engineering}. In the presence of an external
magnetic field, one should replace $\bm{\nabla} \to \bm{\nabla}+ie\mathbf{A}$, where
$\mathbf{A}$ is the electromagnetic vector potential. Following our convention
in Refs.~\cite{engineering,NJL-model1,NJL-model2}, we call $\mathbf{b}_{0}$
the bare chiral shift parameter. As we will see below, the interaction effects renormalize the bare chiral shift parameter
changing the separation between the Weyl nodes to $2\mathbf{b}$, where
$\delta\mathbf{b}=\mathbf{b}-\mathbf{b}_{0}$ is proportional to the magnetic
field $\delta\mathbf{b} \propto \mathbf{B}$ and is determined by the strength
of interaction and the quasiparticle charge density \cite{engineering,NJL-model1,NJL-model2}.

In this study, the interaction Hamiltonian $H_{\rm int}$ is modeled by a local Coulomb interaction,
i.e.,
\begin{equation}
H_{\rm int} = \frac{G_{\rm int}}{2}\int d^3\mathbf{r}\,\psi^{\dagger}(\mathbf{r})\psi(\mathbf{r})
\psi^{\dagger}(\mathbf{r})\psi(\mathbf{r}),
\label{int-Hamiltonian}
\end{equation}
where $G_{\rm int}$ is a dimensionful coupling constant. Before proceeding further with the analysis,
it is convenient to introduce the four-dimensional Dirac matrices in the chiral representation
\begin{equation}
\gamma^0 = \left( \begin{array}{cc} 0 & -I\\ -I & 0 \end{array} \right),\qquad
\bm{\gamma} = \left( \begin{array}{cc} 0& \bm{\sigma} \\  - \bm{\sigma} & 0 \end{array} \right), \qquad \gamma^5 \equiv
i\gamma^0\gamma^1\gamma^2\gamma^3 = \left( \begin{array}{cc} I & 0\\ 0 & -I \end{array} \right),
\label{Dirac-matrices}
\end{equation}
where $I$ is the two-dimensional unit matrix. By making use of these matrices, we can rewrite the Hamiltonian
of the Weyl semimetal in the following simple form:
\begin{equation}
H_{\rm mag} = \int d^3 \mathbf{r}\, \bar{\psi} (\mathbf{r})\left[
-i v_F \left( \bm{\gamma}\cdot(\bm{\nabla}+ie\mathbf{A}) \right)-\mu_{0}\gamma^0
- v_F(\mathbf{b}_{0}\cdot\bm{\gamma})\gamma_5\right]\psi(\mathbf{r})
+\frac{G_{\rm int}}{2}\int d^3\mathbf{r}\,\bar{\psi}(\mathbf{r})\gamma^0\psi(\mathbf{r})\bar{\psi}(\mathbf{r})\gamma^0\psi(\mathbf{r}).
\label{Hamiltonian-Weyl-magnetic}
\end{equation}
For a finite sample of a Weyl semimetal, the above bulk Hamiltonian has to be supplemented by some
boundary conditions on the surface. Let us assume that the sample is an infinite flat slab of macroscopic
thickness $L$. Following the convention of Ref.~\cite{Vishwanath}, we choose the {\em laboratory frame}
of reference so that the sample lies between the boundaries at $y=0$ and $y=L$ in the $y$ direction.
It is infinite in the other two ($x$ and $z$) directions.

To complete the setup of the model, we should specify the boundary conditions for the
quasiparticle wave functions at the surfaces $y=0$ and $y=L$. It is natural to assume that the
quasiparticles outside the Weyl semimetal behave as free particles with the
Dirac mass $m$. This is implemented simply by adding the following
mass term to the Hamiltonian outside the semimetal:
\begin{equation}
H_{m} = \int d^3 \mathbf{r}\,  m   \bar{\psi} (\mathbf{r}) \psi(\mathbf{r}).
\label{Hamiltonian-Dirac-mass}
\end{equation}
Then, the formal description of the system in the whole space can be captured by a single Hamiltonian,
provided that the following conditions are satisfied:
\begin{eqnarray}
\mbox{$y<0$ and $y>L$ (vacuum)}:& \quad &m\neq 0 \quad\mbox{and}\quad \mathbf{b}_{0}= 0,
\label{Delta-b-vacuum}\\
\mbox{$0<y<L$ (semimetal)}: &\quad&m=0 \quad\mbox{and}\quad \mathbf{b}_{0}\neq 0 .
\label{Delta-b-semimetal}
\end{eqnarray}
Since $m$ is much larger than quasiparticle energies in Weyl semimetals (including the work function),
the results obtained with finite $m$ practically cannot be distinguished from those found in the
infinite mass limit $m \rightarrow \infty$. The corresponding boundary conditions at $y=0$ and
$y=L$ in the latter case are known in the studies of graphene as the infinite mass boundary conditions
\cite{Beenakker,Recher}. We will use these boundary conditions in our analysis below.

We would like to note that model (\ref{Hamiltonian-Dirac-mass})--(\ref{Delta-b-semimetal})
is known as the Bogolyubov bag model \cite{Bogoliubov} in relativistic physics. (For a review of bag
models, see Ref.~\cite{bc-symmetry-breaking}.) In these models, hadrons are described as
bags inside which massless quarks are confined. In order to not allow the quarks to leave the
bags, it is assumed that the component of quarks momenta normal to the bag
surface vanishes at the surface. It is interesting that the boundary conditions implied
by this assumption {\it necessarily} break chiral symmetry \cite{bc-symmetry-breaking}.
Since there is spin-flip for quasiparticles associated with the reflection at
the semimetal surface, their chirality changes. This is, in fact, consistent with the topological arguments which require
the existence of topologically stable surface states in Weyl semimetals.
If boundary conditions preserving chiral symmetry were possible, then
taking into account the exact chiral symmetry in the bulk, the surface Fermi
arcs connecting Weyl nodes of opposite chirality would not be allowed as they
obviously break chiral symmetry. Therefore, the impossibility of the existence of
boundary conditions which confine massless Weyl particles in a certain region and
at the same time respect chiral symmetry is intimately connected with the topological
properties of Weyl semimetals.

In the next section, we study the spectrum of the bulk and the Fermi surface
states in semimetals by using the Bogolyubov model (more precisely, its infinite mass
limit version).

\section{Bulk and surface states}
\label{FermiArcs}

Let us derive the spectrum of the low-energy quasiparticle modes in the model of an
infinite flat slab of a Weyl semimetal without the external magnetic field. In this section,
we will ignore the effects of interaction. Later, however, such effects will be taken into
account by replacing the bare value of the chiral shift $\mathbf{b}_{0}$ with its full
counterpart in the interacting theory.

In accordance with the model setup with a Weyl semimetal in the region $0<y<L$
and the vacuum elsewhere, the chiral shift $\mathbf{b}_{0}$ and the Dirac mass
$m$ take different values in different spatial regions; see Eqs.~(\ref{Delta-b-vacuum})
and (\ref{Delta-b-semimetal}). Moreover, in order to describe the most general
case of a Weyl semimetal, we assume that the chiral shift $\mathbf{b}_{0}$ points
in an arbitrary direction and possesses a nonzero component not only in the
direction perpendicular to the slab ($y$ direction), but also in the direction parallel
to the slab. Because of the rotational symmetry in the $xz$ plane, we may choose
without loss of generality a coordinate system, in which the $x$ component of the
chiral shift vanishes, i.e., $\mathbf{b}_{0}=(0, b_{0,y}, b_{0,z})$, where $b_{0,y}$
($b_{0,z}$) is the component of the chiral shift perpendicular (parallel) to the slab.
Then the corresponding one-particle Hamiltonian takes the following general form:
\begin{eqnarray}
{\cal H} &=& v_F\left[\gamma^0(\mathbf{k}\cdot\bm{\gamma})
-i\gamma^1\gamma^2b_{0,z}+i\gamma^1\gamma^3b_{0,y}\right]+\gamma^0 m
\nonumber \\
&=&v_F\left(
\begin{array}{cc} (\bm{\sigma}\cdot\mathbf{k})-\sigma_zb_{0,z}-\sigma_yb_{0,y}  & -m/v_F\\ -m/v_F &
-(\bm{\sigma}\cdot\mathbf{k})-\sigma_zb_{0,z}-\sigma_yb_{0,y}
\end{array} \right) .
\label{Hamiltonian-effective-New}
\end{eqnarray}
The quasiparticle wave function $\Psi(\mathbf{r})$ can be obtained by solving the eigenvalue problem
$({\cal H} -E)\Psi = 0$ in the three separate regions of space: inside the semimetal, $0<y<L$, and in the two vacuum regions outside
it, $y<0$ and $y>L$. When all three solutions are available, the condition of the wave function continuity
at $y=0$ and $y=L$ will have to be imposed. Since there is a symmetry with respect to translations
in the $x$  and $z$ directions, the wave functions are eigenfunctions of the $\hat{k}_x$ and $\hat{k}_z$
components of the operator of momentum in the whole space. Therefore, the problem reduces to finding
the dependence of wave functions on $y$.

Let us start from the analysis in the vacuum regions on each side of the semimetal. In these regions, 
the wave function satisfies the Dirac equation with mass $m$. In order to obtain a normalizable
solution, we will require that the wave functions approach zero at the spatial infinities, $y\to \pm \infty$.
Then, the general solutions in the vacuum regions ($y<0$ and $y>L$) take the following form:
\begin{eqnarray}
\Psi(\mathbf{r})_{y<0} &=& \left(
             \begin{array}{c}
               \phi \\
               \frac{v_F(\sigma_x k_x -i \sigma_y \kappa_{y} +\sigma_z k_z)-E}{m} \phi \\
             \end{array}
           \right)e^{ik_x x+ik_z z +\kappa_{y} y},
\label{App:Psi-y<0a}
\\
\Psi(\mathbf{r})_{y>L}  &=& \left(
             \begin{array}{c}
               \psi \\
               \frac{v_F(\sigma_x k_x  + i \sigma_y \kappa_{y}+\sigma_z k_z)-E}{m} \psi \\
             \end{array}
           \right)e^{ik_x x+ik_z z - (y-L)\kappa_{y}},
\label{App:Psi-y>La}
\end{eqnarray}
where $E=\pm \sqrt{v_F^2(k_x^2+k_z^2-\kappa_{y}^2)+ m^2}$ and $0<\kappa_{y}<\sqrt{(m/v_F)^2+k_x^2+k_z^2}$.
Note that these solutions are given in terms of two-component constant spinors $\phi^{T}=(\phi_1, \phi_2)$ and
$\psi^{T}=(\psi_1, \psi_2)$, whose explicit form will be determined below after matching all pieces of the
wave function at $y=0$ and $y=L$. Due to our assumption discussed in Sec.~\ref{Model},
the vacuum Dirac mass $m$ is much larger than the absolute value of the quasiparticle energy $E$. For the
description of low-energy quasiparticles, this implies that $\kappa_{y} \simeq  m/v_F \gg |k_x|, |k_z|$. Then,
the above vacuum solutions take the following approximate form:
\begin{eqnarray}
\Psi(\mathbf{r})_{y<0} &\simeq & \left(
             \begin{array}{c}
               \phi \\
              -i\sigma_y \phi \\
             \end{array}
           \right)e^{ik_x x+ik_z z+ym/v_F } ,
\label{Psi-y<0}
\\
\Psi(\mathbf{r})_{y>L} &\simeq& \left(
             \begin{array}{c}
               \psi \\
               i\sigma_y \psi \\
             \end{array}
           \right)e^{ik_x x+ik_z z- (y-L)m/v_F} .
\label{Psi-y>L}
\end{eqnarray}
In the bulk of Weyl semimetal ($0<y<L$), the Dirac mass is vanishing and the general form of
the energy eigenfunction solutions can be written in terms of four independent chiral modes, i.e.,
\begin{equation}
\Psi(\mathbf{r})_{0<y<L} = R_1 \Psi^{(R)}_{1}(\mathbf{r})  +R_2 \Psi^{(R)}_{2}(\mathbf{r})
+ L_1 \Psi^{(L)}_{1}(\mathbf{r})  +L_2 \Psi^{(L)}_{2}(\mathbf{r})  ,
\label{Psi-0<y<L}
\end{equation}
where $R_1$, $R_2$, $L_1$, and $L_2$ are weight constants. The explicit form of the right-handed
eigenstates is given by
\begin{eqnarray}
\Psi^{(R)}_{1}(\mathbf{r}) &=& \left(
             \begin{array}{c}
               1\\
               \frac{k_z-b_{0,z}-E/v_F}{ik_y^{\prime}-k_x} \\
               0 \\
               0\\
             \end{array}
           \right)e^{ik_x x+i(b_{0,y}+k_y^{\prime}) y+ik_z z},
\label{Psi-R-1}
\\
\Psi^{(R)}_{2}(\mathbf{r}) &=& \left(
             \begin{array}{c}
               1\\
               \frac{k_z-b_{0,z}-E/v_F}{-ik_y^{\prime}-k_x} \\
               0 \\
               0\\
             \end{array}
           \right)e^{ik_x x+i(b_{0,y}-k_y^{\prime})y+ik_z z},
\label{Psi-R-2}
\end{eqnarray}
with $E=\pm v_F \sqrt{k_x^2+(k_y^{\prime})^2+(k_z - b_{0,z})^2}$,
and the left-handed eigenstates are given by
\begin{eqnarray}
\Psi^{(L)}_{1}(\mathbf{r}) &=& \left(
             \begin{array}{c}
              0 \\
               0 \\
              1\\
               \frac{k_z+b_{0,z}+E/v_F}{ik_y^{\prime \prime}-k_x}   \\
             \end{array}
           \right)e^{ik_x x- i(b_{0,y}-k_y^{\prime \prime})y+ik_z z},
\label{Psi-L-1}
\\
\Psi^{(L)}_{2}(\mathbf{r}) &=& \left(
             \begin{array}{c}
              0 \\
               0 \\
              1\\
               \frac{k_z+b_{0,z}+E/v_F}{-ik_y^{\prime \prime}-k_x}   \\
             \end{array}
           \right)e^{ik_x x-i(b_{0,y}+k_y^{\prime \prime} )y+ik_z z},
\label{Psi-L-2}
\end{eqnarray}
with $E=\pm v_F \sqrt{k_x^2+(k_y^{\prime \prime})^2+(k_z + b_{0,z})^2}$. From the physics
viewpoint, a pair of eigenstates for each chirality describe (bulk) quasiparticles
with the opposite $y$ components of momenta. Since such states have the same
energy, we include both of them in the analysis. Obviously, it suffices to consider only
positive values of $k^{\prime}_y$ and $k^{\prime\prime}_y$; otherwise, double
counting would occur. (For modes that do not propagate in the bulk such as the
Fermi arc modes, the pair of states will correspond to modes with purely imaginary
$k_y^{\prime}$ and $k_y^{\prime \prime} $ localized on the opposite sides of the sample.)
Because all right-handed and left-handed eigenstates in Eq.~(\ref{Psi-0<y<L})
have the same energy, the following constraint must be enforced (without loss of generality,
we assume in what follows that $b_{0,z}>0$):
\begin{equation}
(k_y^{\prime})^2 = (k_y^{\prime \prime})^2+4k_z b_{0,z}.
\label{E-constraint}
\end{equation}
The detailed analysis of the boundary conditions for the quasiparticle wave function
at the two surfaces of the Weyl semimetal at $y=0$ and $y=L$ is performed in
Appendix~\ref{AppModes}. It results in the following spectral equation:
\begin{equation}
\cos\left(2b_{0,y} L \right)+\cos\left(k_y^{\prime}L\right) \cos\left(k_y^{\prime\prime}L\right)
=\frac{4b_{0,z}^2+(k_y^{\prime})^2+(k_y^{\prime\prime})^2}{2 k_y^{\prime} k_y^{\prime\prime}}
\sin\left(k_y^{\prime}L\right) \sin\left(k_y^{\prime\prime} L\right),
\label{quantization-cond2}
\end{equation}
which together with Eq.~(\ref{E-constraint}) determines allowed values of the wave vectors (i.e.,
quantization of $k_y^{\prime}$ and $k_y^{\prime\prime}$). For each choice of the parameters
that satisfies Eqs.~(\ref{E-constraint}) and (\ref{quantization-cond2}),
there exists a well-defined quasiparticle state with the energy given by
$E=\pm v_F \sqrt{k_x^2+(k_y^{\prime})^2+(k_z - b_{0,z})^2}$, or equivalently by
$E=\pm v_F \sqrt{k_x^2+(k_y^{\prime\prime})^2+(k_z + b_{0,z})^2}$.

In order to get an insight into the meaning of the spectral equation (\ref{quantization-cond2}),
let us start by considering the simplest case of a {\em Dirac semimetal} first. In this case $\mathbf{b}_{0}=0$
and the constraint in Eq.~(\ref{E-constraint}) implies that $k_y^{\prime \prime}= k_y^{\prime}
\equiv k_y >0$. Note that, as stated earlier in the general case, the consideration of negative
wave vectors is not needed because that would only result in double counting of the same
chiral eigenstates. By substituting $\mathbf{b}_{0}=0$ into the spectral equation (\ref{quantization-cond2}),
we arrive at the following quantization condition:
\begin{equation}
\cos^2\left(k_y L\right) =0.
\label{quantization-cond-Dirac}
\end{equation}
This is satisfied when the values of momenta $k_y$ take the following discrete values:
$k_{y}(n) =(n+1/2) \pi /L$, where $n$ is a nonnegative integer. Of course, this
momentum quantization is exactly as expected in a finite box of size $L$.

In the case of a Weyl semimetal the analysis of the spectral equation is slightly more
complicated. From the physics viewpoint, this is due to the fact that there can exist at
least two different types of solutions: one of them corresponds to the bulk states, while
the other to the so-called Fermi arc states. Mathematically, as we will see below, the
bulk modes are described by a subset of solutions, in which at least one of the wave vectors
$k_y^{\prime}$ or $k_y^{\prime\prime}$ is real. [While it is necessary that
one of the wave vectors is real, the other may be either real or imaginary, depending
on the outcome of the constraint in Eq.~(\ref{E-constraint}).] The Fermi arc modes,
on the other hand, will correspond to the solutions, in which both wave vectors
$k_y^{\prime}$ and $k_y^{\prime\prime}$ are imaginary.

\subsection{Bulk modes}

We begin our analysis with the {\em bulk modes} by assuming that at least one of the
wave vectors, $k_y^{\prime}$ or $k_y^{\prime\prime}$, is real. Without loss of
generality, let us additionally assume that $k_y^{\prime}$ takes a positive value.
(Recall that considering negative $k_y^{\prime}$ is not needed because it
would only result in double counting of the same solutions.) The other wave vector,
$k_y^{\prime\prime}$, is not independent. It is determined by the constraint in
Eq.~(\ref{E-constraint}); i.e., $k_y^{\prime\prime}=\sqrt{(k_y^{\prime})^2-4b_{0,z} k_z}$.
The actual value of $k_y^{\prime\prime}$ is real when $(k_y^{\prime})^2>4b_{0,z} k_z$
and imaginary otherwise. For the following analysis, however, this is mostly irrelevant. The spectral equation
(\ref{quantization-cond2}) takes the following form for the bulk modes:
\begin{equation}
\cos\left(2b_{0,y} L \right)+\cos(k_y^{\prime} L) \cos\left(L\sqrt{(k_y^{\prime})^2-4b_{0,z} k_z}\right)
-\frac{2b_{0,z}(b_{0,z}-k_z)+(k_y^{\prime})^2}{k_y^{\prime}\sqrt{(k_y^{\prime})^2-4b_{0,z} k_z} }
\sin(k_y^{\prime} L) \sin\left(L\sqrt{(k_y^{\prime})^2-4b_{0,z} k_z}\right)=0.
\label{quantization-cond-bulk}
\end{equation}
In view of the oscillatory function on the left-hand side, this equation admits
numerous solutions. In order to find all of them, one can use numerical methods.
In certain regions of the phase space, however, it is easy to obtain subsets of
approximate solutions by analytical methods. Let us demonstrate this by studying
the two limiting cases of small and large values of $k_y^{\prime}$.

Let us start from the case of small $k_y^{\prime}$ [i.e., $(k_y^{\prime})^2\ll 4|b_{0,z} k_z|$].
In this limit, the spectral equation has the following approximate form:
\begin{equation}
\sin(k_y^{\prime} L) =0.
\label{quantization-cond-small-k1}
\end{equation}
The solutions to this equation are $k_y^{\prime}(n) =n \pi /L$, where $n$ is a
positive integer. It is interesting to note that the quantization of the wave vector in the infrared
region differs from that in the Dirac semimetal; see Eq.~(\ref{quantization-cond-Dirac}) and
the text after it. When measured from the reference point at $b_{0,y}$, the corresponding
values of the wave vectors in a Weyl semimetal are given by integer, rather than half-integer
multiples of $\pi/L$.

In the opposite limit of large $k_y^{\prime}$ [i.e., $(k_y^{\prime})^2\gg 4|b_{0,z} k_z|$],
the approximate spectral equation reads
\begin{equation}
2\cos\left[(k_y^{\prime}+b_{0,y}) L \right]\cos\left[(k_y^{\prime}-b_{0,y}) L \right]=0,
\label{quantization-cond-large-k1}
\end{equation}
and its solutions are given by $k_y^{\prime}(\pm b,n)=\pi (n+1/2)/L \pm b_{0,y}$, where
$n$ is a large positive integer and both choices of the sign are possible. When $b_{0,y}$ is small,
as we see, this quantization of the wave vector reduces to the same as in the Dirac semimetal;
see Eq.~(\ref{quantization-cond-Dirac}) and the text after it.

Each of the bulk states with a given wave vector that satisfies the spectral equation can be
described by a continuous wave function $\Psi(\mathbf{r})$. The explicit form of the
corresponding function is obtained by patching together its vacuum and semimetal parts
in Eqs.~(\ref{Psi-y<0}), (\ref{Psi-y>L}), and (\ref{Psi-0<y<L}). In practice, this can be done
by making use of the results in Appendix~\ref{AppModes}. One can start by determining
the two-component spinor $\phi$ that defines the vacuum solution in the region $y<0$;
see Eq.~(\ref{Psi-y<0}). Up to an overall normalization constant, the components of spinor
$\phi$ are fixed by their ratio in Eq.~(\ref{App:ratio1}). The other vacuum solution (at $y>L$)
is given in terms of spinor $\psi$; see Eq.~(\ref{Psi-y>L}). Both components of spinor $\psi$
can be obtained from $\phi$ by making use of Eqs.~(\ref{App:Psi11}) and (\ref{App:Psi21}).
Finally, the remaining piece of the wave function in the bulk of the semimetal is fixed by adjusting
the values of the weights $R_{1}$, $R_{2}$, $L_{1}$, and $L_{2}$ for each of the four
chiral eigenstates; see Eq.~(\ref{Psi-0<y<L}). The corresponding expressions for the
weights are derived in Eqs.~(\ref{App:R11}) through (\ref{App:L21}). Because of their
cumbersome form, here we will not write down the corresponding explicit expressions for
the wave functions.

\subsection{Surface Fermi arcs}
\label{SurfaceFermiArcs}

Let us now proceed to the case of the {\em Fermi arc modes}. In this case, we assume that both wave vectors,
$k_y^{\prime}$ and $k_y^{\prime\prime}$, are imaginary. It is convenient to introduce
the following shorthand notation: $k_y^{\prime}=i \kappa_1$ and $k_y^{\prime\prime}
=i\kappa_2$, where both $\kappa_1$ and $\kappa_2$ take real values. As before, in order to avoid
double counting, we restrict these parameters to positive values only; i.e., $\kappa_1>0$ and
$\kappa_2>0$. In this case, the spectral relation in Eq.~(\ref{quantization-cond2}) takes the
following form:
\begin{equation}
\cos\left(2b_{0,y} L \right)+\cosh\left( \kappa_1 L \right)\cosh\left(\kappa_2 L\right)
=\frac{4b_{0,z}^2-\kappa_1^2-\kappa_2^2}{2\kappa_1 \kappa_2}\sinh\left( \kappa_1 L \right)\sinh\left(\kappa_2 L\right).
\label{quantization-cond3}
\end{equation}
For a semimetal slab of a macroscopic size $L$, it is generally expected that $|b_{0,z}\pm k_z| L\gg 1$.
In this case, the solutions to the above spectral equation can be obtained analytically. The analysis
is drastically simplified by noting that the resulting solutions will be such that $\kappa_1 L\gg1$ and
$\kappa_2 L\gg1$. Taking this into account, we can rewrite the spectral equation in the following 
approximate form:
\begin{equation}
1\approx \frac{4b_{0,z}^2-\kappa_1^2-\kappa_2^2}{2\kappa_1\kappa_2},
\label{quantization-cond4}
\end{equation}
where exponentially small corrections of order $e^{-(\kappa_1+\kappa_2)L}$ were neglected.
The approximate equation implies that $\kappa_1+\kappa_2 = 2 b_{0,z}$ (when $b_{0,z}>0$). In
accordance with Eq.~(\ref{E-constraint}), the imaginary wave vectors $\kappa_1$ and $\kappa_2$
must also satisfy the constraint $\kappa_2^2-\kappa_1^2=4b_{0,z} k_z$. By solving this set of
two equations, we finally derive the following result:
\begin{eqnarray}
k_y^{\prime}&=& i \kappa_1 = i (b_{0,z} -k_z),
\label{solutions-kappa1}\\
k_y^{\prime\prime}&=& i \kappa_2  =  i(b_{0,z} +k_z).
\label{solutions-kappa2}
\end{eqnarray}
In view of the assumed positivity of the imaginary wave vectors $\kappa_1$ and $\kappa_2$,
the range of validity of this solution is restricted by $|k_z|\leq b_{0,z}$. As is easy to see, this
solution describes the Fermi arc modes with the dispersion relations given by
$E= \pm v_F\sqrt{k_x^2+(k_y^{\prime})^2+(b_{0,z}-k_z)^2}=\pm v_F k_x $.

By making use of the results in Appendix~\ref{AppModes}, we can easily construct the
wave functions of the Fermi arc modes. For the branch of modes propagating in the
$+x$ direction, the energy is given by $E=v_F k_x$. In this case, as follows from
Eq.~(\ref{App:ratio1}), the ratio of the two components of spinor $\phi$ is approximately
equal to $1$. Furthermore, from Eqs.~(\ref{App:R11}) through (\ref{App:L21}), one
also finds that $R_1 = -L_1 = \phi_1$ and $R_2 =L_2 =0$. Therefore, the explicit form
of the wave function in the region of semimetal $0<y<L$ up to an overall normalization constant is given by
\begin{eqnarray}
\Psi(\mathbf{r})_{0<y<L} &\simeq& \left(
             \begin{array}{c}
               1\\
               1 \\
               0 \\
               0\\
             \end{array}
           \right)e^{ik_x x+ib_{0,y} y+ik_z z-(b_{0,z}-k_z)y}
+\left(
             \begin{array}{c}
              0 \\
               0 \\
              -1\\
              1   \\
             \end{array}
           \right)e^{ik_x x-ib_{0,y} y+ik_z z-(b_{0,z}+k_z)y}.
\end{eqnarray}
As we see, for negative values of the wave vector $k_z$ (recall that $|k_z|<b_{0,z}$), the exponent
$e^{-(b_{0,z}+k_z)y}$ in the second term falls off slower with increasing $y$. Therefore, the corresponding
left-handed spinor dominates the wave function. Similarly, for positive values of $k_z$, the dominant
part of the wave function is given by the first term, which is a right-handed spinor. In both cases,
it is clear that the peak of the absolute value of the wave function is located at $y=0$. Therefore,
we conclude that the arc modes with the dispersion relation $E=v_F k_x$ are localized at $y=0$
boundary of the semimetal.

Similarly, one can show that the Fermi arc modes with the dispersion relation $E=-v_F k_x$, i.e.,
modes propagating in the $-x$ direction, have their wave functions localized near the $y=L$
boundary of the semimetal. This is indeed obvious from the explicit form of the wave functions,
whose derivation we omit here and present only the final results. For negative values $k_z$,
the Fermi arc modes with $E=-v_F k_x$ are described by the following left-handed spinor:
\begin{eqnarray}
\Psi(\mathbf{r})_{0<y<L} &\simeq& \left(
             \begin{array}{c}
              0 \\
               0 \\
              1\\
              1  \\             \end{array}
           \right)e^{ik_x x-ib_{0,y} y+ik_z z+(b_{0,z}+k_z)(y-L)},
\end{eqnarray}
while, for positive values of $k_z$, the Fermi arc modes are described
by the following right-handed spinor:
\begin{eqnarray}
\Psi(\mathbf{r})_{0<y<L} &\simeq & \left(
             \begin{array}{c}
               1\\
               -1 \\
               0 \\
               0\\
             \end{array}
           \right)e^{ik_x x+ib_{0,y} y+ik_z z+(b_{0,z}-k_z)(y-L)} .
\end{eqnarray}
Before finishing this section, let us summarize the key observations about the effect of the chiral shift
on the spectrum of Fermi arc modes. There exist two branches of the Fermi arc modes that are
localized on the opposite surfaces of the semimetal (i.e., at $y=0$ and $y=L$). The ``lengths" of
the arcs are determined exclusively by the value of the chiral shift component $b_{0,z}$ parallel to 
the surface; i.e., $|k_z|<b_{0,z}$. The dispersion relations of the two branches of Fermi arc modes,
$E=\pm v_F k_x$, depend only on the wave vector $k_x$, which is also parallel to the surface
of the semimetal, but perpendicular to the chiral shift. The modes on the opposite sides of the
semimetal move in the opposite directions along the $x$ axis. At the same time, the chirality 
of the Fermi arc modes evolves from being predominantly left-handed at $k_z=-b_{0,z}$ to 
predominantly right-handed at $k_z=b_{0,z}$.

\section{Interaction effects on the chiral shift in a magnetic field}
\label{Interaction}

In this section, we study the effect of quasiparticle interactions on the chiral shift in a Weyl 
semimetal in the presence of a magnetic field. This is a generalization of the previous study 
\cite{engineering} in Dirac semimetals, where the bare chiral shift was absent. In general, 
the direction of the magnetic field can be arbitrary. As shown in Ref.~\cite{engineering},
interactions dynamically generate a contribution to the chiral shift $\delta \mathbf{b}$ 
proportional to the magnetic filed and pointing in the same direction; i.e., 
$\delta \mathbf{b} \propto \mathbf{B}$. Because of this, below we will allow the most 
general orientation of the chiral shift. 

While studying the dynamical generation of the chiral shift, it is convenient to rotate 
the reference frame so that the magnetic field points in the $+z$ direction.
Let us start by deriving the gap equation for the fermion propagator in the Weyl semimetal 
with the Hamiltonian (\ref{Hamiltonian-Weyl-magnetic}). The formal expression for the 
inverse free fermion propagator reads
\begin{equation}
iS^{-1}(r,r^\prime) = \left[(i\partial_t+\mu_{0})\gamma^0
-v_F(\bm{\pi}\cdot\bm{\gamma})+v_F(\mathbf{b}_{0}\cdot\bm{\gamma})\gamma^5\right]\delta^{4}(r-r^\prime),
\label{sinverse}
\end{equation}
where $r=(t,\mathbf{r})$ and $\bm{\pi} \equiv -i \bm{\nabla} + e\mathbf{A}$ is the canonical 
momentum. [When the explicit form is needed, we will use the vector potential in the Landau 
gauge, $\mathbf{A}= (0, x B,0)$, which describes an external magnetic field of strength $B$ 
that points in the $+z$ direction.] Similarly, the inverse full fermion propagator in the normal 
phase is given by
\begin{equation}
iG^{-1}(r,r^\prime)= \Big[(i\partial_t+\mu )\gamma^0 - v_F(\bm{\pi}\cdot\bm{\gamma})
+v_F (\mathbf{b} \cdot \bm{\gamma})\gamma^5\Big]\delta^{4}(r- r^\prime),
\label{ginverse}
\end{equation}
where $\mathbf{b}$ is a renormalized chiral shift that may differ from the bare one, $\mathbf{b}_{0}$.
In general, the dynamical parameter $\mu$ in the full propagator may also differ from its bare counterpart
$\mu_{0}$.

In this study, we use the mean-field approximation at weak coupling. Instead 
of the perturbative analysis, however, we utilize the formalism of the Baym-Kadanoff (BK) 
effective action \cite{BK-CJT}, which leads to a self-consistent Schwinger--Dyson equation 
for the quasiparticle propagator. The main advantage of the BK formalism is its ability 
to capture nonperturbative effects such as the spontaneous symmetry breaking, which 
may develop even at arbitrarily weak coupling. This is in contrast to the perturbative 
analysis which misses all such effects. To leading order in coupling, the BK effective action 
\cite{BK-CJT} reads
\begin{equation}
\Gamma(G) = -i\,\mathrm{Tr}\left[\mathrm{Ln}G^{-1}+S^{-1}G -1\right] -\frac{G_{\rm int}}{2}\int d^4r \Big(
\mbox{tr}{\big[\gamma^0G(r,r)\big]}\mbox{tr}{\big[\gamma^0G(r,r)\big]} -\mbox{tr}{\big[\gamma^0G(r,r)\gamma^0G(r,r)\big]} \Big),
\label{BK-CJT-LR-1}
\end{equation}
where the trace $\mathrm{Tr}[\ldots]$ in the first term is taken in functional sense, while the trace
$\mathrm{tr}[\ldots]$ in the other terms operates only in spinor space. The Schwinger--Dyson
equation for the full propagator follows from the condition of extremum of the effective action
$\delta \Gamma(G)/\delta G =0$. In the problem at hand, this gives the following equation:
\begin{equation}
iG^{-1}(r,r^\prime) = iS^{-1}(r,r^\prime) - G_{\rm int} \left\{ \gamma^0 G(r,r) \gamma^0
- \gamma^0\, \mbox{tr}[\gamma^0G(r,r)]\right\}\delta^{4}(r-r^{\prime}).
\label{gap}
\end{equation}
The derivation of the fermion propagator in the Landau-level representation is given in
Appendix~\ref{AppPropagator}. In general, in such a representation all dynamical functions
(e.g., $\mu$ and $\mathbf{b}$) have an explicit dependence on the Landau-level index $n$.
In the case of the contact interaction that is used in this study, however, the corresponding
dynamical parameters are independent of $n$ that drastically simplifies the analysis.

By making use of the propagator derived in Appendix~\ref{AppPropagator}, we obtain the following
explicit form of the equations for $\mu$ and $\mathbf{b}$:
\begin{eqnarray}
\hspace*{-5mm}
\mu-\mu_0 &=& \frac{3G_{\rm int}}{4}\int_{-\infty}^{\infty}\frac{d\omega}{2\pi}
\int_{-\infty}^{\infty}\frac{d^3\mathbf{q}}{(2\pi)^3}
\mathrm{tr}\left\{\gamma^0 \left[
  \bar{G}_{||}(\omega, q_z, \mathbf{q}_{\perp}+\mathbf{b}_{\perp}){\cal P}_{+}
+\bar{G}_{||}(\omega, q_z, \mathbf{q}_{\perp}-\mathbf{b}_{\perp}){\cal P}_{-}
\right]\right\},
\label{gap-equation-trace-mu}
\\
\hspace*{-5mm}
b_x-b_{0,x} &=&  -\frac{G_{\rm int}}{4v_F}\int_{-\infty}^{\infty}\frac{d\omega}{2\pi}
\int_{-\infty}^{\infty}\frac{d^3\mathbf{q}}{(2\pi)^3}
\mathrm{tr}\left\{\gamma^1\gamma^5\left[
   \bar{G}_{||}(\omega, q_z, \mathbf{q}_{\perp}+\mathbf{b}_{\perp}){\cal P}_{+}
+ \bar{G}_{||}(\omega, q_z, \mathbf{q}_{\perp}-\mathbf{b}_{\perp}){\cal P}_{-}
\right]\right\},
\label{gap-equation-trace-b-x}
\\
\hspace*{-5mm}
b_y-b_{0,y} &=&  -\frac{G_{\rm int}}{4v_F}\int_{-\infty}^{\infty}\frac{d\omega}{2\pi} \int_{-\infty}^{\infty}\frac{d^3\mathbf{q}}{(2\pi)^3}
\mathrm{tr}\left\{\gamma^2\gamma^5\left[
\bar{G}_{||}(\omega, q_z, \mathbf{q}_{\perp}+\mathbf{b}_{\perp}){\cal P}_{+}
+\bar{G}_{||}(\omega, q_z, \mathbf{q}_{\perp}-\mathbf{b}_{\perp}){\cal P}_{-}
\right]\right\},
\label{gap-equation-trace-b-y}
\\
\hspace*{-5mm}
b_z-b_{0,z} &=& - \frac{G_{\rm int}}{4v_F}\int_{-\infty}^{\infty}\frac{d\omega}{2\pi}
\int_{-\infty}^{\infty}\frac{d^3\mathbf{q}}{(2\pi)^3}
\mathrm{tr}\left\{\gamma^3\gamma^5\left[
  \bar{G}_{||}(\omega, q_z, \mathbf{q}_{\perp}+\mathbf{b}_{\perp}){\cal P}_{+}
+\bar{G}_{||}(\omega, q_z, \mathbf{q}_{\perp}-\mathbf{b}_{\perp}){\cal P}_{-}
\right]\right\},
\label{gap-equation-trace-b-z}
\end{eqnarray}
where, by definition, ${\cal P}_{\pm} = (1\pm \gamma^5)/2$ are the chirality projectors and
$\bar{G}_{||}(\omega,q_z, \mathbf{q}_{\perp}\pm\mathbf{b}_{\perp})$ is the Fourier transform of the translation
invariant part of the full fermion propagator, in which the chiral shift has only a nonzero
component along the magnetic field, i.e., $b_{z}$. By making use of the integrals in Appendix~\ref{AppIntegrals}
and performing some simplifications, this system of gap equations reduces down to a set
of algebraic equations, i.e.,
\begin{eqnarray}
\mu-\mu_0 &=&-\frac{3G_{\rm int} \left(\mu-s_{\perp}v_F b_z\right) }{8v_F\pi^2 l^2}
-\frac{3G_{\rm int} \sign{(\mu)}}{4v_F\pi^2 l^2}
\sum_{n=1}^{N_{\rm max}^{(\mu)} } \sqrt{\mu^2-2nv_F^2|eB|} ,
\label{Int-mu-Normal-Sys}
\\
b_z-b_{0,z} &=& -\frac{G_{\rm int} s_{\perp}\left(\mu-s_{\perp}v_F b_z\right) }{8v_F^2\pi^2 l^2}
+ \sum_{n=1}^{\infty} \frac{G_{\rm int} b_z}{4v_F\pi^2 l^2},
\label{Int-b-z-Normal-Sys}
\end{eqnarray}
together with $b_x=b_{0,x}$ and $b_y=b_{0,y}$, which come as a result of integrating odd functions
over the perpendicular components $\tilde{\mathbf{q}}_{\perp} \equiv \mathbf{q}_{\perp}\pm\mathbf{b}_{\perp}$.
[Strictly speaking, there exist radiative corrections independent of the magnetic field to 
all components of the bare shift, $b_{0,x}$, $b_{0,y}$, and $b_{0,z}$ \cite{Abanin}. For the purposes 
of this study, the corresponding corrections are absorbed into the renormalization of $\mathbf{b}_{0}$.
Then, the correction $\delta\mathbf{b}$ contains only the magnetic field dependent part.]
In the last two equations, we introduced the magnetic length 
$l\equiv 1/\sqrt{|eB|}$, as well as the following shorthand notation: $s_{\perp}=\mbox{sign}(eB)$ 
and $N_{\rm max}^{(\mu)} =\left[ \mu^2/(2v_F^2|eB|)\right]$, where $\left[\ldots\right]$ denotes 
the integer part.

Let us now derive the solutions to the above set of the gap equations. We start by noting that while
the sum in Eq.~(\ref{Int-mu-Normal-Sys}) is finite, the sum in Eq.~(\ref{Int-b-z-Normal-Sys}) is divergent.
The latter is an artifact of using a nonrenormalizable model with a contact interaction \cite{NJL-model1,NJL-model2}.
In the situation at hand, this difficulty can be avoided by introducing a cutoff at
$N_{\rm max}^{(\Lambda)}=\left[ \Lambda^2/|eB| \right]$,
where $\Lambda=\pi/a$ is a wave vector cutoff related to the lattice spacing $a$.
Then, the  system of equations (\ref{Int-mu-Normal-Sys})--(\ref{Int-b-z-Normal-Sys}) can
be rewritten as follows:
\begin{eqnarray}
\mu-\mu_0 &=& -3 g \left(\mu+\frac{ g  \mu -s_{\perp}v_F b_{0,z}}{1- g -2 g N_{\rm max}^{(\Lambda)}} +2 g
\sign{(\mu)}\sum_{n=1}^{N_{\rm max}^{(\mu)}}\sqrt{\mu^2-2nv_F^2|eB|} \right),
\label{delta-mu}
\\
\delta{b} &\equiv& b_z - b_{0,z}
= -\frac{ g  \left[s_{\perp}\mu - v_F b_{0,z} \left(1+2N_{\rm max}^{(\Lambda)}\right)\right]}{v_F\left(1- g -2 g
N_{\rm max}^{(\Lambda)}\right)},
\label{delta-z}
\end{eqnarray}
where $ g  = G_{\rm int}/(8\pi^2v_F l^2)$ is a dimensionless coupling.
The first equation is an implicit expression for $\mu$ and can be easily solved
in the general case by numerical methods. In a special case with $ g  \ll 1$ and $\mu_0^2<2v_F^2|eB|$, it admits
an approximate analytical solution
\begin{equation}
\label{Int-mu-Normal-Sys-App}
\mu \approx  \frac{\mu_0+3s_{\perp} g v_F b_{0,z} }{1+3  g }.
\end{equation}
At sufficiently weak coupling, we find that there is a weak dependence of $\mu$ and $b_z$
on the magnetic field. We will use the above result for $\delta{b}$ in the analysis 
of the quantum oscillations of the density of states involving Fermi arcs in the next section.

\section{Quantum oscillations of the density of states involving Fermi arcs}
\label{oscillations}

\begin{figure}[!ht]
\begin{center}
\includegraphics[width=0.45\textwidth]{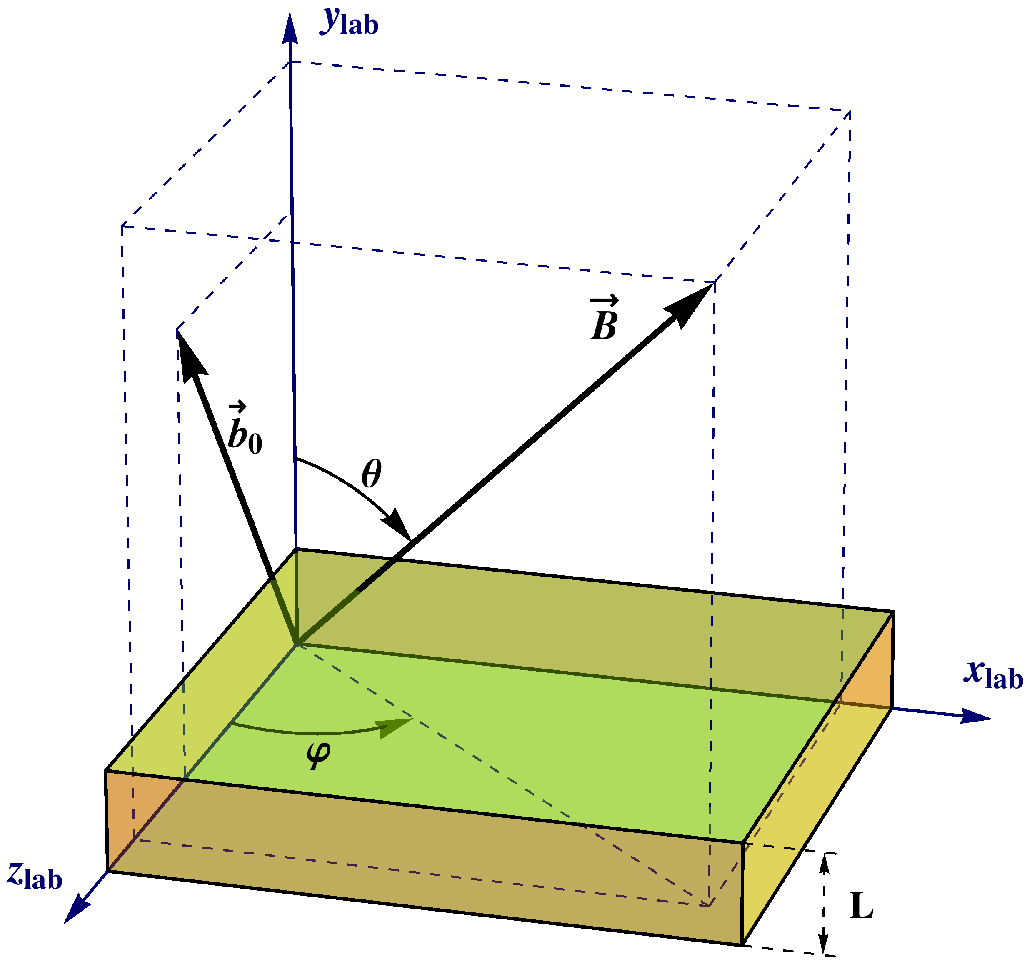}\hspace{0.075\textwidth}
\includegraphics[width=0.45\textwidth]{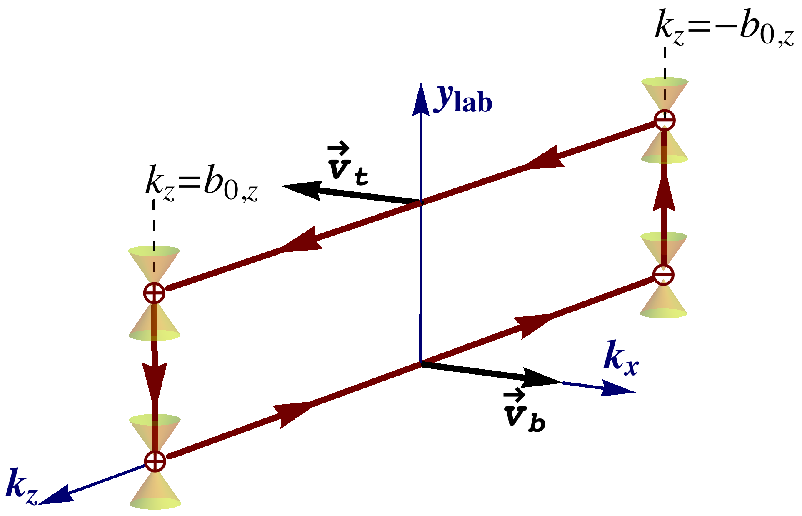}
\caption{(Color online) The model setup and the choice of the laboratory reference frame
used in the current study (left panel) and a schematic representation of the closed
quasiparticle orbits in a magnetic field involving the Fermi arcs (right panel).}
\label{fig:ref-frames}
\end{center}
\end{figure}

In this section we will study the influence of interactions on the oscillation of the density of states associated
with the Fermi arc modes in the Weyl semimetals. We will start by first reproducing the results obtained
in Ref.~\cite{Vishwanath} and then study the influence of interactions on the period of oscillations. In the
{\em laboratory frame} of reference, see Fig.~\ref{fig:ref-frames}, the Weyl semimetal slab is characterized by the
Weyl nodes located at $\pm\mathbf{b}_{0}=\pm(0, b_{0,y}, b_{0,z})$, where $b_{0,y}$ ($b_{0,z}$) is the component of the
chiral shift perpendicular (parallel) to the slab. As our analysis in Sec.~\ref{FermiArcs} revealed,
the quasiparticle velocity of the Fermi arc modes is $\mathbf{v}_{\rm b}=(v_F, 0, 0)$
on the bottom surface ($y=0$) and is $\mathbf{v}_{\rm t}=(-v_F, 0, 0)$ on the top surface ($y=L$).
The direction of the magnetic field $\mathbf{B}=B\, \hat{\mathbf{n}}$ is specified by the unit
vector $\hat{\mathbf{n}} = (\sin\theta\sin\varphi,\cos\theta,\sin\theta\cos\varphi)$, where $\theta$ is
the angle between the $y_{\rm lab}$ axis and the direction of the magnetic field, and $\varphi$ is the angle between
the $z_{\rm lab}$ axis and the projection of $\mathbf{B}$ onto the $z_{\rm lab}x_{\rm lab}$ plane,
see the left panel in Fig.~\ref{fig:ref-frames}.

The Fermi arc states are localized on the surfaces of the semimetal and are characterized by
wave vectors $k_x$ and $k_z$. Since the velocities of these modes are parallel to the $x$ axis,
only the $z$ component of the quasiclassical equation of motion in the magnetic field is
nontrivial; i.e.,
\begin{eqnarray}
&\partial_t k_{z} = -e\left[\mathbf{v}_{\rm b}\times\mathbf{B}\right]_{z}=-e v_F B_y , &\qquad \mbox{(bottom surface)},
\label{k_z-bottom}\\
&\partial_t k_{z} = -e\left[\mathbf{v}_{\rm t}\times\mathbf{B}\right]_{z}=e v_F B_y , &\qquad \mbox{(top surface)}.
\label{k_z-top}
\label{arc-momentum}
\end{eqnarray}
Therefore, the corresponding quasiparticles slide along the bottom (top) Fermi arc from the right-handed
(left-handed) node at $k_z=b_{0,z}$ ($k_z=-b_{0,z}$) to the left-handed (right-handed) node at $k_z=-b_{0,z}$
($k_z=b_{0,z}$), where $b_{0,z}$ is the component of the chiral shift parallel to the surface of the
semimetal. The surface parts of the quasiparticle orbits are connected with each other via the gapless
bulk modes of fixed chirality. The corresponding closed orbits are schematically shown in Fig.~\ref{fig:ref-frames},
where we use a mixed coordinate--wave vector representation combining the out-of-plane $y$ axis with the in-plane
$k_x$  and $k_z$ axes.

The semiclassical quantization condition for the closed orbits involving the Fermi arc modes reads
\cite{Vishwanath}
\begin{equation}
E_n t = 2\pi(n+\gamma),
\label{semiclassical-condition}
\end{equation}
where $n$ is an integer and $\gamma$ is an unknown phase shift that can be determined only via the
rigorous quantum analysis. The total time $t$ includes the time of quasiparticle propagation through the bulk 
$t_{\rm bulk}=2 L/(v_F \cos{\theta})$, where the presence of $\cos{\theta}$ is related to the fact that the 
movement through the bulk occurs only along the magnetic field, and the time it takes for the surface 
Fermi arcs to evolve from one chirality
node to the other. The latter can be estimated from Eqs.~(\ref{k_z-bottom}) and (\ref{k_z-top}), i.e.,
$t_{\rm arcs}=2k_0/(v_F e B_y)$, where $k_0$ is the arc length in the reciprocal space. As we
showed in Sec.~\ref{SurfaceFermiArcs}, it is determined by the component of the (full) chiral shift
parallel to the surface of the semimetal. Therefore, Eq.~(\ref{semiclassical-condition}) implies that
\begin{equation}
E_n = \frac{\pi v_F(n+\gamma)}{L/\cos{\theta}+k_0/(e B_y)}.
\label{semiclassical-condition-01}
\end{equation}
If the Fermi energy $\mu$ is fixed and the magnetic field is varied, the density of states, associated
with the corresponding close quasiclassical orbits, will be oscillating. The peaks of such oscillations
occur when the Fermi energy crosses the energy levels in Eq.~(\ref{semiclassical-condition-01}).
Taking into account that $B_y = B\cos\theta$, we derive the following discrete values of the magnetic
field that correspond to the maxima of the density of state oscillations:
\begin{equation}
\frac{1}{B_n} = \frac{e}{k_0} \left(\frac{\pi v_F \cos{\theta}}{\mu}(n+\gamma)-L\right).
\label{magnetic-inverse-g0-1}
\end{equation}
By noting that the expression on the right-hand side should remain
positive definite, we conclude that the smallest possible value of $n$ is given by 
$n_{\rm min}=\left[ \mu L/(\pi v_F \cos{\theta})-\gamma+1\right]$, where $[\ldots]$ denotes the
integer part. This corresponds to the saturation value of the magnetic field
$B_{\rm sat} \equiv B_{n_{\rm min}}$, above which no more oscillations will be
observed.

In order to quantify the effects of interaction in the Weyl semimetal in a magnetic
field, we will study how the density of states, involving the Fermi arc modes,
oscillates as a function of the inverse magnetic field. In general, as the above
analysis shows, the period of oscillations is given by
\begin{equation}
T_{1/B}=\frac{e\pi v_F \cos{\theta}}{\mu k_0}.
\label{period-oscillations}
\end{equation}

\subsection{Interaction effects}

\begin{figure*}[!ht]
\includegraphics[width=0.45\textwidth]{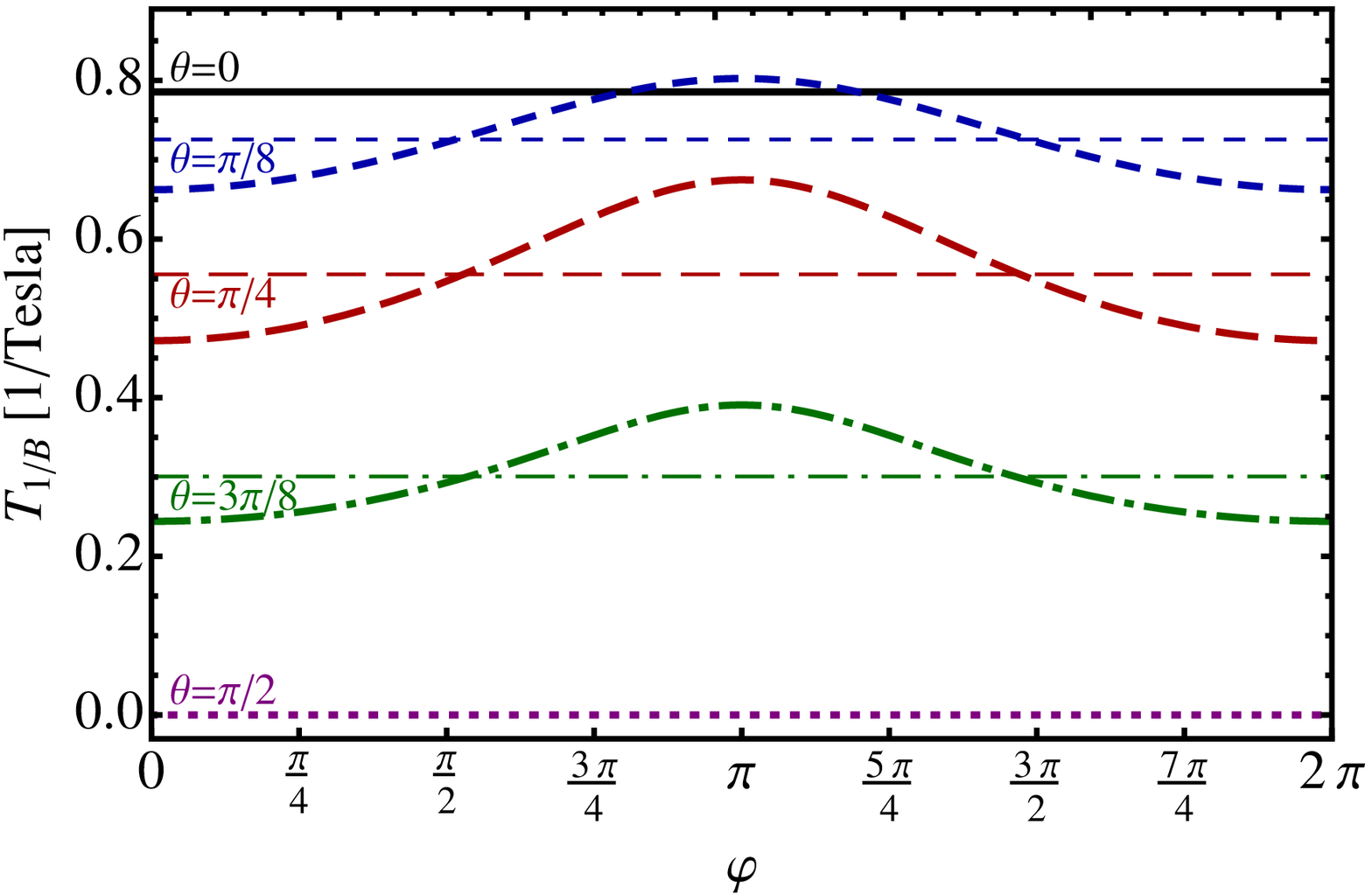}\hspace{0.05\textwidth}
\includegraphics[width=0.45\textwidth]{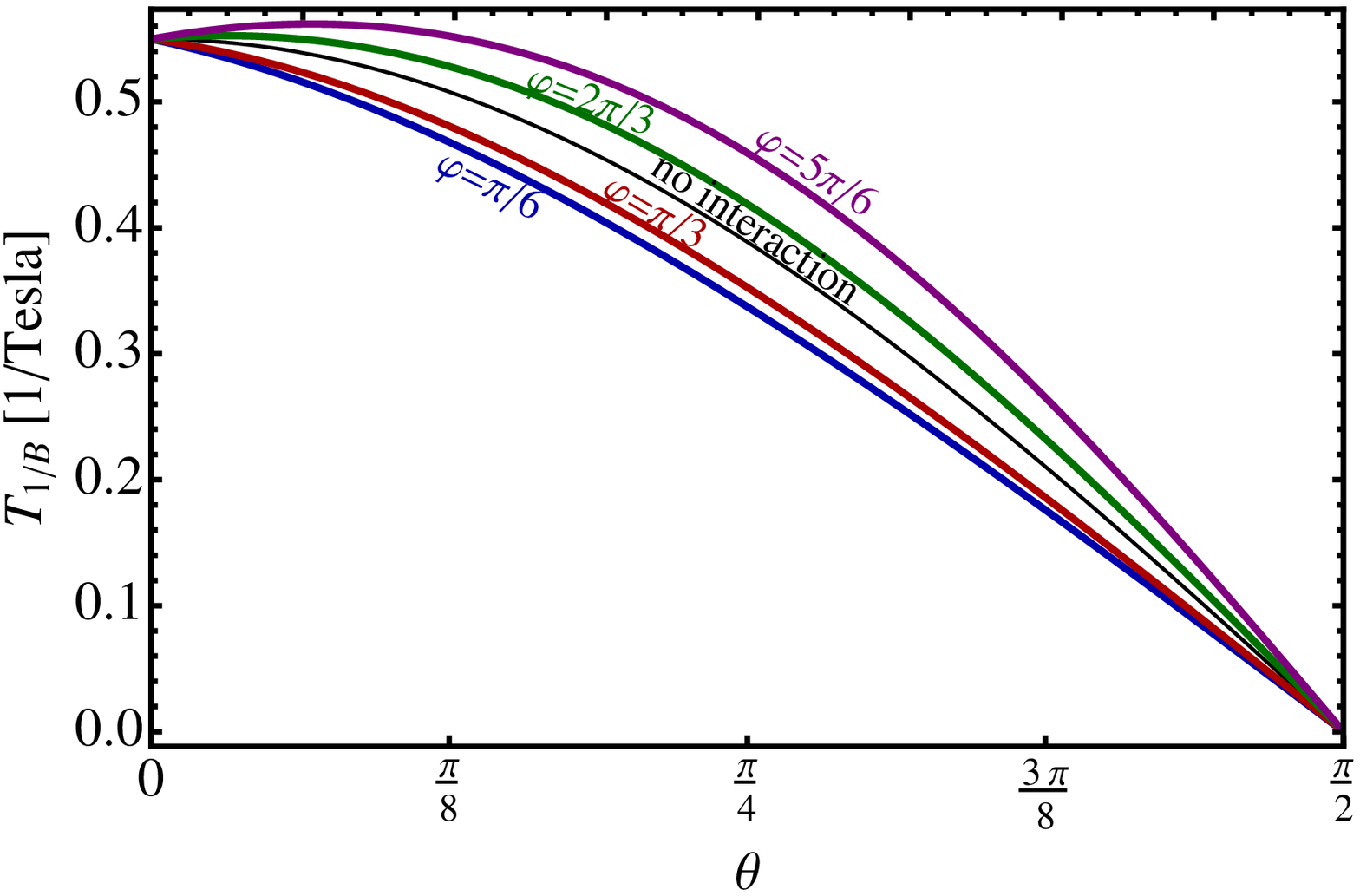}
\caption{(Color online)
The numerical results for the period of the density of state oscillations as a function of
angle $\varphi$ (left panel) and angle $\theta$ (right panel) for several fixed values of
the other angle. The results with and without interaction effects are shown by thick and
thin lines, respectively.}
\label{fig:period}
\end{figure*}

The interaction effects on the chiral shift were analyzed in Sec.~\ref{Interaction}
and can be summarized as follows. The full chiral shift takes the following form:
$\mathbf{b}=\mathbf{b}_{0}+\delta{b}\, \hat{\mathbf{n}}$, where $\mathbf{b}_{0}$ is the
chiral shift in absence of the magnetic field, $\delta{b}$ is the magnitude of the correction
to the chiral shift given by Eq.~(\ref{delta-z}), and $\hat{\mathbf{n}}$ is the unit vector pointing
in the direction of the field. Therefore, in the {\em laboratory frame} of reference, in which
the $x$ component of the bare chiral shift vanishes, $b_{0,x} =0 $, the explicit expressions
for the components of the full chiral shift in a Weyl semimetal in a magnetic field
read
\begin{eqnarray}
b_{x} &=&  \delta{b} \sin\theta\sin\varphi,
\label{dynamical-b-x}\\
b_{y} &=& b_{0,y} + \delta{b} \cos\theta,
\label{dynamical-b-y}  \\
b_{z} &=& b_{0,z} +\delta{b} \sin\theta\cos\varphi.
\label{coordinate-connection-3}
\end{eqnarray}
Since the length of the Fermi arcs is determined by the component of the chiral shift
{\em parallel} to the surface (the analysis in Sec.~\ref{SurfaceFermiArcs} can be easily
generalized to the case where the $x$ component of chiral shift is nonvanishing), we
include the effects of interaction in Eq.~(\ref{period-oscillations}) by replacing the bare
arc length with its interaction modified expression; i.e.,
$k_0\to 2 b_{\parallel}=2\sqrt{b_{x}^2+b_{z}^2}$. Then by making use of
Eqs.~(\ref{dynamical-b-x}) and (\ref{coordinate-connection-3}), we obtain our final
result for the period of oscillations in the interacting case with an arbitrary oriented
magnetic field
\begin{equation}
T_{1/B}=\frac{e\pi v_F \cos{\theta}}{2\mu\sqrt{\left(b_{0,z}\right)^2 + 2b_{0,z} \delta{b} \cos\varphi  \sin\theta
+\left(\delta{b} \sin\theta \right)^2}}.
\label{ArcLengthInteraction}
\end{equation}
Since the Fermi arc length $2b_{||}=2\sqrt{\left(b_{0,z}\right)^2 + 2b_{0,z} \delta{b} \cos\varphi  \sin\theta
+\left(\delta{b} \sin\theta \right)^2}$ depends on the magnetic field, strictly
speaking, the period will be drifting with the varying magnetic field. The numerical results for
the period as a function of angle $\theta$ ($\varphi$) are shown in the left (right) panels of
Fig.~\ref{fig:period} for several fixed values of angle $\varphi$ ($\theta$). [Note that we display only
the results for $\theta<\pi/2$ because $T_{1/B}(\pi-\theta)=T_{1/B}(\theta)$.] The thick (thin) lines
represent the results with (without) taking interaction into account. To obtain these results we
fixed the model parameters as follows:
$b_{0,z}=10^{8}\, \mbox{m}^{-1}$,
$L=1.5\times 10^{-7}\, \mbox{m}$,
$a=0.5 \,\mbox{nm}$,
$v_F=5\times 10^{5}\,  \mbox{m/s}$,
$\mu=10\, \mbox{meV}$. In order to extract the qualitative effects in the cleanest form,
let use a moderately strong coupling, $g =1/(10\Lambda^2 l^2)$. While such a set of model
parameters is representative, it does not correspond to any specific material.

As seen from the numerical results in Fig.~\ref{fig:period}, the most important qualitative
effect is the $\varphi$ dependence of the oscillation period, which appears only in
the case with interaction. The emergence of such a dependence is easy
to understand from the physics viewpoint. Indeed, in the noninteracting theory, the
chiral shift, as well as the length of the Fermi arcs $k_0$ associated with it, are
independent of the magnetic field. The situation drastically changes in the interacting
theory when a correction to the chiral shift $\delta\mathbf{b}$ parallel to $\mathbf{B}$
is generated. As follows from Eq.~(\ref{ArcLengthInteraction}), such a correction
introduces a nontrivial dependence of the Fermi arc length on $\varphi$ as soon as
the component of the magnetic field parallel to the surface is nonzero (i.e., $\theta\neq 0$).
At $\theta = 0$, the period of quantum oscillations does not depend on $\phi$ because
in this case the magnetic field is perpendicular to the surface of a semimetal and, therefore,
$\mathbf{b}_{||}=\mathbf{b}_{0,||}$. Also, as one can see on the left panel of Fig.~\ref{fig:period}
and in Eq.~(\ref{ArcLengthInteraction}), the period of quantum oscillations vanishes
at $\theta=\pi/2$. In this case the component of the magnetic field perpendicular to the
surface is absent and, therefore, the quasiclassical motion along the Fermi arcs is forbidden.
The maximum value of the peak of the period of the oscillations as a function of $\varphi$
takes place at 
\begin{equation}
\theta=\frac{1}{2}\arccos\left(\frac{(\delta{b}/b_{0,z})^2}{2 -(\delta{b}/b_{0,z})^2}\right) .
\end{equation}
Measuring this angle in experiment would make it possible to determine the value of the ratio 
$\delta{b}/b_{0,z}$ and, thus, quantify the interaction effects in Weyl semimetals. However, 
a complete fit of the angular dependence in Eq.~(\ref{ArcLengthInteraction}) to the period
obtained in experiment will allow us to extract not only $\delta{b}/b_{0,z}$, but also the value
of  $b_{0,z}$ if the chemical potential and the Fermi velocity are known.

\section{Superposition of surface and bulk oscillations}
\label{SurfaceBulk}

As suggested in Ref.~\cite{Vishwanath}, the quantum oscillation of the density of states involving
the Fermi arcs can be most visible in surface-sensitive probes. This is a sensible proposal because the
weight of the surface arcs is generally expected to be small compared to the bulk modes. It should
be noted, however, that the quantitative estimate of the corresponding suppression is lacking.
Moreover, one could also argue that the relative contribution of the surface arcs into the total
density of states may become observable if the thickness of the semimetal sample is sufficiently
small. In order to explore this issue quantitatively, in this section we model the superposition
of surface and bulk oscillations.

Let us start from presenting the expression for the known density of the bulk modes near one of the
Weyl nodes multiplied by the total number of nodes $N_W$ \cite{BulkMagOsc},
\begin{equation}
N(\omega)_{\rm bulk}
= \frac{N_W|eB|}{4\pi^2 v_F} \left(1+2\sum_{n=1}^{n_{\rm max}} 
\frac{|\omega|}{\sqrt{\omega^2- 2n v_F^2|eB|}} \right),
\label{A-DOS-Carbotte-1}
\end{equation}
where $n_{\rm max}=\left[\omega^2 /(2v_F^2|eB|)\right]$. The oscillatory part of this density of states 
can be made more realistic by including nonzero quasiparticle widths. Usually, this can be modeled 
simply by replacing $\delta$-function 
contributions in the quasiparticle spectral function with the Lorentz distribution of a nonzero width
$\Gamma$. However, in the case of bulk states, this is not the best approach because the Lorentz 
distribution with a constant parameter $\Gamma$ produces a divergent Landau-level sum in the 
final expression. Instead, following the approach of Ref.~\cite{Gusynin}, here we include a nonzero 
width in the final expression for the oscillatory part of the density of states [cf. Eq.~(13) in 
Ref.~\cite{BulkMagOsc}],
\begin{equation}
N(\omega)_{\rm bulk}^{\Gamma} = \frac{\omega^2N_W}{2\pi^2 v_F^3} + \frac{\omega\,N_W}{\pi^2 v_F^2} 
\sum_{k=1}^{\infty} e^{-\frac{2\pi k \omega \Gamma}{v_F^2|eB|}} \sqrt{\frac{|eB| }{2k}} \left[
   \cos{\left(\frac{\pi \omega^2 k }{v_F^2|eB|}\right)} C\left(\frac{\omega\sqrt{2k}}{v_F\sqrt{|eB|}}\right) 
+  \sin{\left(\frac{\pi \omega^2 k }{v_F^2|eB|}\right)} S\left(\frac{\omega\sqrt{2k}}{v_F\sqrt{|eB|}}\right)\right],
\label{A-DOS-Carbotte-2}
\end{equation}
where $\exp\left(-\frac{2\pi k \omega \Gamma}{v_F^2|eB|}\right)$ is the Dingle factor, $C(z)$ and $S(z)$ are the Fresnel cosine and sine integrals, respectively.

\begin{figure*}[!ht]
\includegraphics[width=0.32\textwidth]{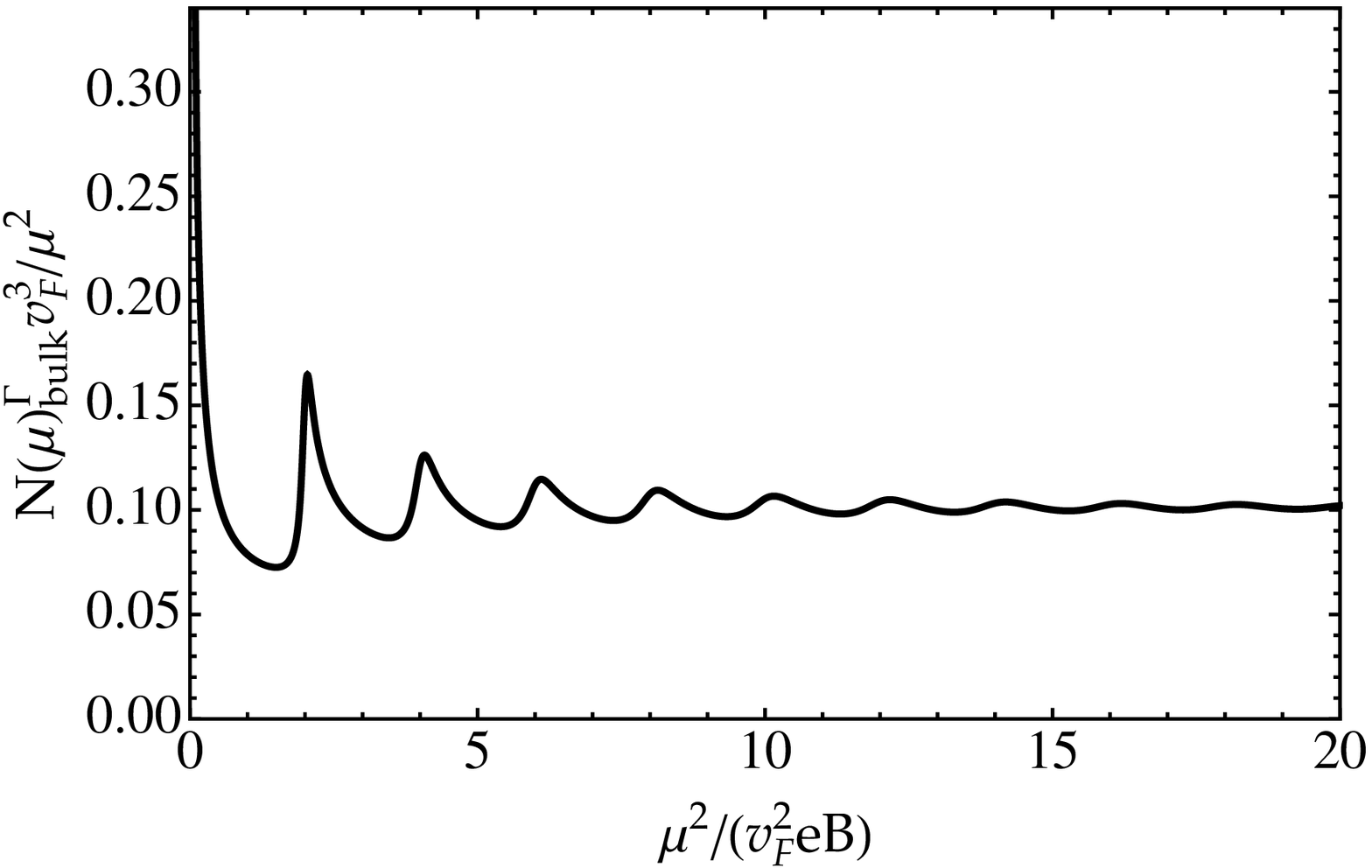}
\includegraphics[width=0.32\textwidth]{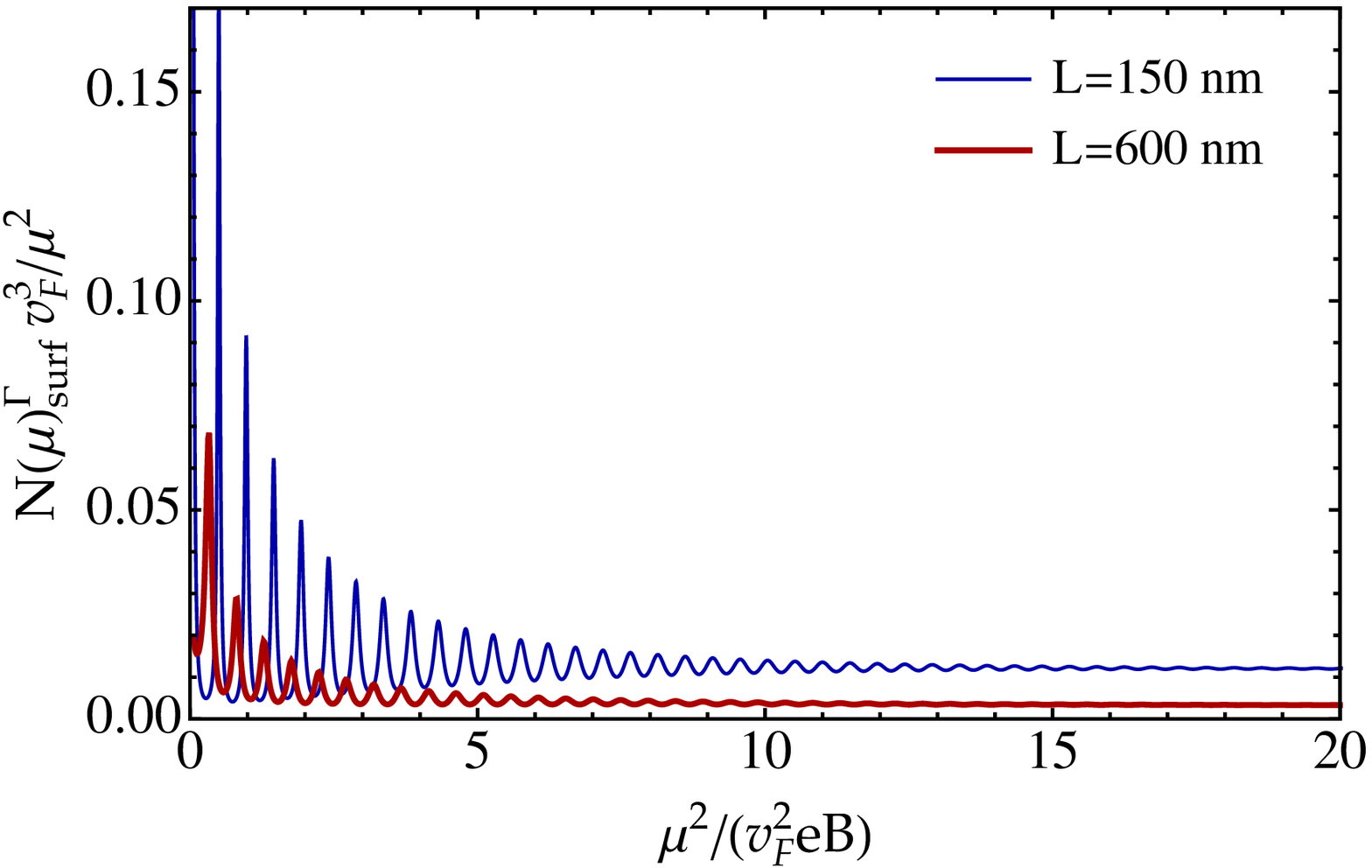}
\includegraphics[width=0.32\textwidth]{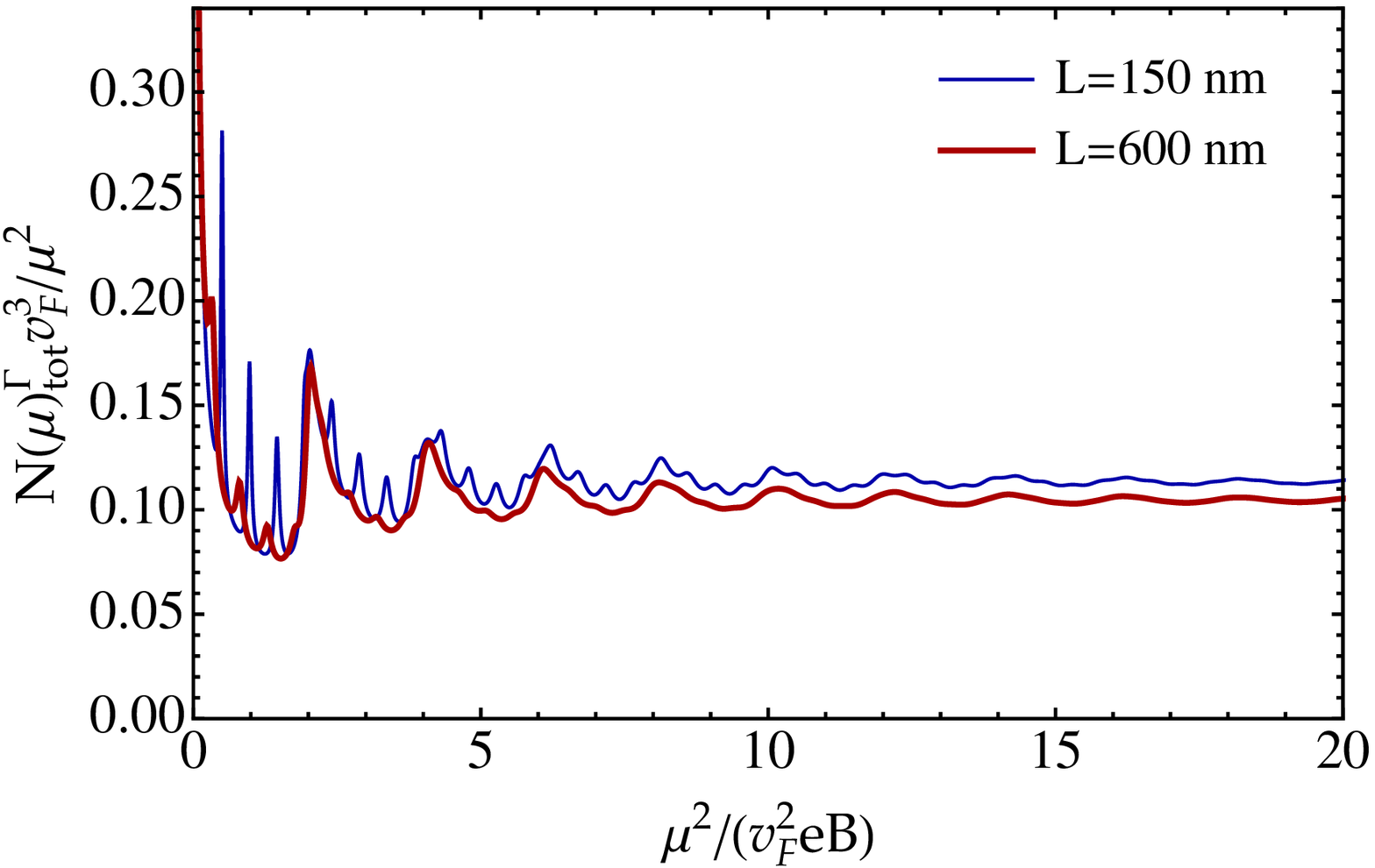}
\caption{(Color online) The density of states for bulk modes alone (left panel), the 
quasiparticle modes involving surface arc states (middle panel), and the 
superposition of both types of modes (right panel). The results are for the 
quasiparticle width parameter $\Gamma=0.2\, \mbox{meV}$.}
\label{fig:Bulk+SurfaceGamma_0.2meV}
\end{figure*}

The density of states involving the surface Fermi arc modes can be estimated by the following
formal expression:
\begin{equation}
N(\omega)_{\rm surf} = \frac{|eB|}{2\pi^2 L}\sum_{n=n_{\rm min}}^{\infty} \left[ \delta(\omega-E_n)+\delta(\omega+E_n) \right] ,
\label{A-Surface}
\end{equation}
where the explicit expression for the energy $E_n$ is given in Eq.~(\ref{semiclassical-condition-01}).
(For brevity, we call such states ``surface states" below, but they truly are surface-bulk mixed
states associated with pairs of Fermi arcs and bulk states connecting them.) It should be noted 
that the expression is proportional to $1/L$, which is a geometric factor connected to the fact 
that the Fermi arc states live only on the surface. In order to make a more realistic model of 
quasiparticles with nonvanishing spectral widths, we will replace the $\delta$ functions in 
Eq.~(\ref{A-Surface}) by the Lorentz distribution with a constant width parameter $\Gamma$;
i.e.,
\begin{equation}
N(\omega)_{\rm surf}^{\Gamma}=  \frac{|eB| }{2\pi^3  L}  \sum_{n=n_{\rm min}}^{\infty}
\left(\frac{\Gamma}{(E_n-\omega)^2+\Gamma^2} + \frac{\Gamma}{(E_n+\omega)^2+\Gamma^2} \right).
\label{A-Surface-11}
\end{equation}
(One may also use the Gaussian distribution to model the effects of a nonzero quasiparticle 
width. We checked, however, that the final numerical results for the oscillations of density 
of states will remain qualitatively the same.) The corresponding sum can be performed 
analytically and the result reads
\begin{equation}
N(\omega)_{\rm surf}^{\Gamma} = 
\frac{|eB| (L+k_0/|eB|)}{2\pi^4 v_F L \cos\theta} \mbox{Im}\left[ 
\psi\left( n_{\rm min}+\gamma+ \frac{(\omega+i\Gamma)(L+k_0/|eB|)}{\pi v_F \cos\theta}
\right)
+\psi\left(n_{\rm min}+\gamma- \frac{(\omega-i\Gamma)(L+k_0/|eB|)}{\pi v_F \cos\theta}
\right)
\right] .
 \label{A-Surface-Exact}
\end{equation}
The numerical results for the bulk and surface densities of states at the Fermi surface, 
as well as for their superposition, are presented in Fig.~\ref{fig:Bulk+SurfaceGamma_0.2meV}. 
To plot the results, we used the quasiparticle width $\Gamma=0.2\, \mbox{meV}$ and the same 
model parameters as in the previous section (i.e., 
$b_{0,z}=10^{8}\, \mbox{m}^{-1}$,
$v_F=5\times 10^{5}\,  \mbox{m/s}$,
$\mu=10\, \mbox{meV}$), 
except for the semimetal thickness $L$, which was varied in order to understand the 
dependence of the oscillations on the sample size. In the figure, we show the results 
for two different values of the thickness: $L=1.5\times 10^{-7}\, \mbox{m}$ (blue lines) 
and $L=6\times 10^{-7}\, \mbox{m}$ (red lines). As expected, the density of the bulk 
states does not depend on $L$, but the contribution of the surface states dramatically 
decreases with $L$. Indeed, from the model results in Fig.~\ref{fig:Bulk+SurfaceGamma_0.2meV}, 
we see that the superposition of the bulk and surface types of oscillations is quite 
well resolved for a relatively thin sample with $L=1.5\times 10^{-7}\, \mbox{m}$ (blue lines), 
but not so much for a sample with $L=6\times 10^{-7}\, \mbox{m}$ (red lines). 
We also see from Fig.~\ref{fig:Bulk+SurfaceGamma_0.2meV} that the oscillations 
are well resolved for the relatively small value of the quasiparticle width chosen. 
We checked, however, that the oscillations quickly become overdamped when $\Gamma$ 
increases. For example, when $\Gamma=0.5\, \mbox{meV}$, it is already hard 
to resolve any features in the interplay of the two types of contributions beyond 
the second peak of the bulk oscillations.

In summary, while the surface states give a small contribution in general, under a favorable 
choice of the parameters, they may nevertheless be observable in the bulk probes. The 
necessary conditions are likely to be (i) a relatively small semimetal thickness, 
(ii) a high-quality clean semimetal, and (iii) a sufficiently low temperature. 

\section{Discussion}
\label{Discussion}

In the current study, we introduced a new model setup to describe the quasiparticle states
in a Weyl semimetal with a slab geometry of finite thickness. We used a Bogolyubov model
\cite{Bogoliubov} for the boundary conditions at the semimetal surfaces. Such a model was
originally used in relativistic physics for the description of hadrons within the framework of
the so-called bag models of quarks \cite{bc-symmetry-breaking}. From the physics viewpoint, the
model assumes that the quasiparticles have an infinite Dirac mass in the vacuum regions
outside the semimetal. Interestingly, while such a treatment has some serious limitations
in the context of quarks and hadrons \cite{bc-symmetry-breaking}, it appears to be completely
justified and physical in the context of semimetals.

Within this framework, we derived analytically the spectral equation for quasiparticle states
in a Weyl semimetal with the most general orientation of the chiral shift. Also, by making use
of this equation, we then analyzed the spectrum of the surface Fermi arcs, as well as the
bulk modes. As expected, our qualitative results for the Fermi arcs agree with those in
Ref.~\cite{Vishwanath}, where a different set of boundary conditions was utilized and the
chiral shift was restricted to be parallel to the semimetal surface.

We also investigated the effect of interaction on the oscillation of the density of states
involving the surface Fermi arc modes. We found that the interaction induces a
small, but nonzero correction to the chiral shift, which points in the same direction as
the magnetic field. As a result, the period of the oscillations of the density of states
acquires a nonnegligible dependence on the angle $\varphi$ between the directions
of the bare chiral shift and the magnetic field projected onto the plane of the semimetal
boundary. The corresponding angular dependence is present for a generic
direction of the magnetic field, but vanishes in the two special cases when the magnetic
field is either parallel or perpendicular to the surface. In the first case the quasiparticle
motion along the surface arc is frozen, while in the second case the length of the arc
is not affected by interactions. Our results suggest that the experimental study of the
corresponding angular dependence of the period could provide a quantitative measure
of the chiral shift $\mathbf{b}_{0}$ and its modification $\delta\mathbf{b}$ due the 
interaction effects.

It is appropriate to mention that the most natural approach to observe the quantum 
oscillation of the density of states involving the Fermi arcs is to use surface-sensitive 
probes \cite{Vishwanath}. However, our study here suggests that this type of 
oscillation may be also observable in bulk probes if the thickness of the semimetal 
sample is not too large. Then, the interplay of the surface and bulk oscillations may 
be resolved if the Weyl semimetal is of high quality and the measurements are done
at sufficiently low temperatures. Both of these requirements are necessary to ensure  
a small quasiparticle width, which generally has a destructive effect on the oscillations 
as well as their interplay. Despite their limitations, the potential ability of the 
bulk measurements to shed light on the Fermi arcs can be very important for several 
reasons: This may provide an independent way of testing the theoretical predictions, 
allow the use of alternative (perhaps, easier) experimental techniques, and provide the 
tool to explore a potentially interesting dependence of the topological effects on the 
semimetal thickness.

At last, but not least, we would like to emphasize that the present analysis also implies
that even Dirac semimetals, where the bare chiral shift vanishes, $\mathbf{b}_{0}=0$,
will necessarily get a nonzero chiral shift in the the presence of a magnetic field after
interactions are taken into account (for earlier studies, see also Refs.~\cite{engineering,
NJL-model1,NJL-model2}). This may suggest that the oscillations of the density of states
associated with the quasiparticles orbits involving Fermi arcs could also be observable
in magnetized Dirac semimetals such as Bi$_{1-x}$Sb$_x$ (with $x \approx 0.03$),
A$_3$Bi (with A=Na, K, or Rb), and Cd$_3$As$_2$ \cite{Sun,Fang}. If the present
analysis is extrapolated to this limiting case, the corresponding period of oscillations
$T_{1/B}$ will be expressed in terms of the interaction induced length of the Fermi
arc $2b_{||}$. Since the value of interaction-induced
correction $\delta{b}$ to the bare chiral shift is likely to be small, one may
expect that the period $T_{1/B}$ will be generally much larger than in Weyl
semimetals with a nonzero $\mathbf{b}_{0}$. On top of that, the period will have the
following very strong functional dependence on the orientation of the magnetic
field: $T_{1/B}\propto \cot\theta$.

\acknowledgments
This work was finished when V.A.M. visited the Institute for Theoretical Physics, 
Goethe University, Frankfurt am Main. He expresses his gratitude to the Helmholz 
International Center ``HIC for FAIR" at Goethe University for financial support. 
He also express his gratitude to Prof. Dirk Rischke for useful discussions and 
his warm hospitality. The work of E.V.G. was supported partially by SFFR of 
Ukraine Grant No.~F53.2/028, the European FP7 program Grant No. SIMTECH 
246937, and the grant STCU No. 5716-2 ``Development of Graphene Technologies 
and Investigation of Graphene-based Nanostructures for Nanoelectronics and 
Optoelectronics." The work of V.A.M. was supported by the Natural Sciences 
and Engineering Research Council of Canada. The work of I.A.S. was supported 
in part by the U.S. National Science Foundation under Grants No.~PHY-0969844 
and No.~PHY-1404232.

\appendix

\section{Derivation of the spectral equation in a Weyl semimetal with boundary}
\label{AppModes}

In this appendix, we present the detailed derivation of the spectral equation
(\ref{quantization-cond2}). It is obtained by imposing the condition of the
wave function continuity at the $y=0$ and $y=L$ surfaces of the semimetal.

By matching the vacuum piece of the wave function (\ref{Psi-y<0}) with the semimetal
one (\ref{Psi-0<y<L}) at $y=0$ boundary, we obtain the following constraint:
\begin{eqnarray}
\left(
\begin{array}{c}
 \phi_1 \\
 \phi_2 \\
 -\phi_2 \\
 \phi_1 \\
 \end{array}
\right) = \left(
\begin{array}{c}
 R_1 +R_2 \\
 \frac{k_z-b_{0,z}-E/v_F}{ik_y^{\prime}-k_x}R_1 -\frac{k_z-b_{0,z}-E/v_F}{ik_y^{\prime}+k_x} R_2  \\
 L_1+L_2 \\
 \frac{k_z+b_{0,z}+E/v_F}{ik_y^{\prime\prime}-k_x}L_1-\frac{k_z+b_{0,z}+E/v_F}{ik_y^{\prime \prime}+k_x}  L_2  \\
 \end{array}
\right) ,
\label{App:condition-y=0}
\end{eqnarray}
which allows us to determine the weight of each chiral component in the general solution for the semimetal wave function; i.e.,
\begin{eqnarray}
2k_y^{\prime}R_1 &=& (k_y^{\prime}+i k_x) \phi_1  + i (b_{0,z}-k_z-E/v_F) \phi_2,
\label{App:R11}\\
2k_y^{\prime}R_2 &=&  (k_y^{\prime}-i k_x)  \phi_1  - i  (b_{0,z}-k_z-E/v_F) \phi_2,
\label{App:R21}\\
2k_y^{\prime\prime}L_1 &=&  -(k_y^{\prime\prime}+i k_x) \phi_2  - i (b_{0,z}+k_z-E/v_F) \phi_1,
\label{App:L11}\\
2k_y^{\prime\prime}L_2 &=& -(k_y^{\prime\prime}-i k_x) \phi_2  + i  (b_{0,z}+k_z-E/v_F) \phi_1.
\label{App:L21}
\end{eqnarray}
Similarly, by matching the vacuum solution (\ref{Psi-y>L}) with the solution (\ref{Psi-0<y<L}) at $y=L$ boundary,
we obtain the following constraint:
\begin{eqnarray}
\left(
\begin{array}{c}
 R_1 e^{i(k_y^{\prime}+b_{0,y})L} +R_2 e^{i(b_{0,y}-k_y^{\prime}) L}\\
 \frac{k_z-b_{0,z}-E/v_F}{ik_y^{\prime}-k_x}R_1 e^{i(k_y^{\prime}+b_{0,y})L}
 -\frac{k_z-b_{0,z}-E/v_F}{ik_y^{\prime}+k_x} R_2 e^{i(b_{0,y}-k_y^{\prime}) L} \\
 L_1 e^{i(k_y^{\prime\prime}-b_{0,y})L} +L_2 e^{-i(b_{0,y}+k_y^{\prime\prime}) L} \\
 \frac{k_z+b_{0,z}+E/v_F}{ik_y^{\prime \prime}-k_x }L_1 e^{i(k_y^{\prime\prime}-b_{0,y})L}
 -\frac{k_z+b_{0,z}+E/v_F}{ik_y^{\prime \prime}+k_x }  L_2  e^{-i(b_{0,y}+k_y^{\prime\prime}) L} \\
 \end{array}
\right) =
\left(
\begin{array}{c}
 \psi_1 \\
 \psi_2 \\
 \psi_2 \\
 -\psi_1 \\
 \end{array}
\right).
\label{App:condition-y=L}
\end{eqnarray}
This is satisfied when the weights of the chiral components in the general solution are given by
\begin{eqnarray}
2k_y^{\prime}R_1 e^{i(k_y^{\prime}+b_{0,y})L}&=&
(k_y^{\prime}+i k_x) \psi_1  + i (b_{0,z}-k_z-E/v_F) \psi_2,
\label{App:R12}\\
2k_y^{\prime}R_2 e^{i(b_{0,y}-k_y^{\prime}) L}&=&
(k_y^{\prime}-i k_x)  \psi_1  - i (b_{0,z}-k_z-E/v_F) \psi_2,
\label{App:R22}\\
2k_y^{\prime\prime}L_1 e^{i(k_y^{\prime\prime}-b_{0,y})L}&=&
(k_y^{\prime\prime}+i k_x)  \psi_2  + i (b_{0,z}+k_z-E/v_F) \psi_1,
\label{App:L12}\\
2k_y^{\prime\prime}L_2 e^{-i(b_{0,y}+k_y^{\prime\prime}) L} &=&
(k_y^{\prime\prime}-i k_x) \psi_2  - i (b_{0,z}+k_z-E/v_F) \psi_1.
\label{App:L22}
\end{eqnarray}
By matching the two sets of definitions for the coefficients $R_1$ and $R_2$,
see Eqs.~(\ref{App:R11}), (\ref{App:R21}), (\ref{App:R12}), and (\ref{App:R22}),
we are able to express the spinor components $\psi_1$ and $\psi_2$ in terms of $\phi_1$ and $\phi_2$;
i.e.,
\begin{eqnarray}
\psi_1 e^{- ib_{0,y} L}&=& \cos\left(k_y^{\prime}L\right) \phi_1
-\left(\frac{k_x}{k_y^{\prime}}\phi_1+\frac{b_{0,z}-k_z-E/v_F}{k_y^{\prime}}\phi_2
\right)\sin\left(k_y^{\prime}L\right)
\label{App:Psi11}
,\\
\psi_2 e^{- ib_{0,y} L}&=&\cos\left(k_y^{\prime}L\right) \phi_2
+\left(\frac{k_x}{k_y^{\prime}}\phi_2
-\frac{b_{0,z}-k_z+E/v_F}{k_y^{\prime}}\phi_1
\right)\sin\left(k_y^{\prime}L\right).
\label{App:Psi21}
\end{eqnarray}
The remaining two sets of definitions for the coefficients $L_1$ and $L_2$, see Eqs.~(\ref{App:L11}), (\ref{App:L21}),
(\ref{App:L12}), and (\ref{App:L22}), lead to the two additional constraints; i.e.,
\begin{eqnarray}
\psi_1 e^{ib_{0,y} L}&=& - \cos\left(k_y^{\prime\prime}L\right) \phi_1
-\left(\frac{k_x}{k_y^{\prime\prime}}\phi_1-\frac{b_{0,z}+k_z+E/v_F}{k_y^{\prime\prime}}\phi_2
\right)\sin\left(k_y^{\prime\prime}L\right) ,
\label{App:Psi12}
\\
\psi_2 e^{ib_{0,y} L}&=& - \cos\left(k_y^{\prime\prime}L\right) \phi_2
+\left(\frac{k_x}{k_y^{\prime\prime}}\phi_2
+\frac{b_{0,z}+k_z-E/v_F}{k_y^{\prime\prime}}\phi_1
\right)\sin\left(k_y^{\prime\prime}L\right).
\label{App:Psi22}
\end{eqnarray}
By identifying the two different expressions for $\psi_1$ in Eqs.~(\ref{App:Psi11}) and (\ref{App:Psi12}), we arrive at the
following ratio of the spinor components $\phi_2$ and $\phi_1$:
\begin{equation}
\frac{\phi_2}{\phi_1} = \frac{F_1-F_2}{G_1-G_2}.
\label{App:ratio1}
\end{equation}
Similarly, by identifying the two different expressions for $\psi_2$ in Eqs.~(\ref{App:Psi21}) and (\ref{App:Psi22}),
we arrive at yet another expression for same ratio,
\begin{equation}
\frac{\phi_2}{\phi_1} = \frac{G_1+G_2}{F_1+F_2},
\label{App:ratio2}
\end{equation}
where
\begin{eqnarray}
F_1 &=& - k_y^{\prime}k_y^{\prime\prime}
\left[e^{ib_{0,y} L}\cos\left(k_y^{\prime}L\right)
+e^{-ib_{0,y} L}\cos\left(k_y^{\prime\prime}L\right) \right],\\
F_2 &=& k_x\left[-e^{ib_{0,y} L}k_y^{\prime\prime}\sin\left(k_y^{\prime}L\right)
+e^{-ib_{0,y} L}k_y^{\prime}\sin\left(k_y^{\prime\prime}L\right)\right],\\
G_1 &=& -e^{ib_{0,y} L}k_y^{\prime\prime}(b_{0,z}-k_z)\sin\left(k_y^{\prime}L\right)
-e^{-ib_{0,y} L}k_y^{\prime}(b_{0,z}+k_z)\sin\left(k_y^{\prime\prime}L\right),\\
G_2 &=& E/v_F\left[-e^{ib_{0,y} L}k_y^{\prime\prime}\sin\left(k_y^{\prime}L\right)
+e^{-ib_{0,y} L}k_y^{\prime}\sin\left(k_y^{\prime\prime}L\right)\right].
\end{eqnarray}
Notice that $F_2 =  (v_Fk_x/E) G_2$. Finally, the consistency of the two expressions in
Eqs.~(\ref{App:ratio1}) and (\ref{App:ratio2}) leads to the following spectral equation:
\begin{equation}
F_1^2 - F_2^2 = G_1^2 - G_2^2,
\end{equation}
which, after some simplifications, reduces down to the following algebraic relation:
\begin{equation}
2(k_y^{\prime})^2(k_y^{\prime\prime})^2\left[
\cos\left(2b_{0,y} L \right)+\cos\left(k_y^{\prime}L\right) \cos\left(k_y^{\prime\prime}L\right)
\right] =k_y^{\prime}k_y^{\prime\prime}
\left[4b_{0,z}^2+(k_y^{\prime})^2+(k_y^{\prime\prime})^2\right]
\sin\left(k_y^{\prime}L\right) \sin\left(k_y^{\prime\prime}L\right).
\label{App:spectral-eq}
\end{equation}
To proceed with the analysis, let us first show that there are no nontrivial wave functions for either
of the solutions $k_y^{\prime}=0$ or $k_y^{\prime\prime}=0$. Indeed, let us assume to the contrary
that, for example, $k_y^{\prime}=0$ is a valid solution. Then, as follows from Eqs.~(\ref{App:R11}) and (\ref{App:R21}),
the condition of having finite values of $R_1$ and $R_2$ would require that $\phi_1=\phi_2=0$.
From the constraint in Eq.~(\ref{App:condition-y=0}), moreover, we would find that $R_1=-R_2$ and
$L_1=L_2=0$. At $k_y^{\prime}=0$, this is a trivial wave function, however.  Similarly, imposing
$k_y^{\prime\prime}=0$ produces only a trivial solution.

Without loss of generality, therefore, we assume that $k_y^{\prime}\neq 0$ and
$k_y^{\prime\prime}\neq 0$. Then, after dividing both sides of the
spectral equation (\ref{App:spectral-eq}) by $(k_y^{\prime})^2(k_y^{\prime\prime})^2$,
we arrive at the final form of the spectral equation in Eq.~(\ref{quantization-cond2}) in the main text.

\section{Fermion propagator in a magnetic field}
\label{AppPropagator}

In this appendix, we present the explicit form of the fermion propagator in a magnetic field,
which is used in the analysis of the gap equation in Sec.~\ref{Interaction}. For our purposes,
we need to consider the case of vanishing Dirac mass, but the most general orientation of the
chiral shift $\mathbf{b}=(b_x,b_y,b_z)$. The corresponding result can be easily obtained by
making use of the known fermion propagator in the special case when the chiral shift
$\mathbf{b}=(0,0,b_z)$ is parallel to a magnetic field $\mathbf{B}=(0,0,B)$ \cite{NJL-model1,NJL-model2}.
In the notation of the present study, the explicit form of the corresponding propagator reads
\begin{equation}
G_{\parallel}(\omega,k_{z};\mathbf{r}_{\perp}, \mathbf{r}_{\perp}^{\prime})=e^{i\Phi(\mathbf{r}_{\perp}, \mathbf{r}_{\perp}^{\prime})}
\bar{G}_{\parallel}(\omega,k_{z};\mathbf{r}_{\perp}- \mathbf{r}_{\perp}^{\prime}),
\label{G-b-parallel-B}
\end{equation}
where $\Phi(\mathbf{r}_{\perp}, \mathbf{r}_{\perp}^{\prime})=- s_{\perp}(x+x^{\prime})(y-y^{\prime})/(2l^2)$
is the Schwinger phase, $s_{\perp}=\mbox{sign}(eB)$, and $l\equiv 1/\sqrt{|eB|}$ is the magnetic length.
The Fourier transform of the translation invariant part of the propagator is given by
\begin{equation}
\bar{G}_{\parallel}(\omega,k_{z},\mathbf{k}_{\perp})
= \int d^2\mathbf{r}_{\perp}\, e^{-i (\mathbf{k}_{\perp} \cdot \mathbf{r}_{\perp})}\bar{G}(\omega,k_{z};\mathbf{r}_{\perp})
= ie^{-k_{\perp}^2 l^{2}}\sum_{n=0}^{\infty} \sum_{\chi=\pm}
\frac{ (-1)^n  D^{(\chi)}_{n}(\omega,k^{3},\mathbf{k}_{\perp})  {\cal P}_{\chi}}{(\omega+\mu)^2 - v_F^2 (k_{z}+\chi b_z)^2-2nv_F^2|e B |},
\label{GDn-new}
\end{equation}
where $\mathbf{k}_{\perp}=(k_x,k_y)$ is the ``transverse momentum"
and the $n$th Landau level contribution is determined by
\begin{equation}
D^{(\chi)}_{n}(\omega,k^{3},\mathbf{k}_{\perp}) = 2\left[(\omega+\mu )\gamma^0 -v_F \gamma^3(k_{z}+\chi b_z) \right]\left[P_{-1}L_n\left(2 k_{\perp}^2 l^{2}\right)
-P_{+1}L_{n-1}\left(2 k_{\perp}^2 l^{2}\right)\right]
 + 4v_F(\mathbf{k}_{\perp}\cdot\bm{\gamma}_{\perp}) L_{n-1}^1\left(2 k_{\perp}^2 l^{2}\right).
\label{Dn}
\end{equation}
Here  $P_{\pm} =\left[1 \pm  i\,s_{\perp} \gamma^1 \gamma^2 \right]/2$ are the spin projection operators
and $L_{n}^{\alpha}(z)$ are the generalized Laguerre polynomials \cite{Gradshtein}.
By definition, $L_{n} (z)=L_{n}^{0} (z)$ and $L_{-1}^\alpha \equiv 0$.

In order to derive the expression for the fermion propagator with the most general chiral shift,
$\mathbf{b}=(b_x,b_y,b_z)$, we note that its formal definition
\begin{eqnarray}
G(\omega,k_z;\mathbf{r}_{\perp}, \mathbf{r}_{\perp}^{\prime})
&=& \langle \mathbf{r}_{\perp}|
\left[ (\omega+\mu )\gamma^0 - v_F(\bm{\pi}_{\perp}\cdot\bm{\gamma}_{\perp})
- v_F \gamma^3 (k_z + b_z\gamma^5) - v_F (\mathbf{b}_{\perp}\cdot\bm{\gamma}_{\perp})\gamma^5 \right]^{-1}
 | \mathbf{r}_{\perp}^{\prime}\rangle \nonumber\\
&=& \langle \mathbf{r}_{\perp}|
e^{i (\mathbf{b}_{\perp}\cdot\mathbf{r}_{\perp}) \gamma^5}
\left[ (\omega+\mu )\gamma^0 - v_F(\bm{\pi}_{\perp}\cdot\bm{\gamma}_{\perp})
- v_F \gamma^3 (k_z + b_z\gamma^5)   \right]^{-1}
e^{i (\mathbf{b}_{\perp}\cdot\mathbf{r}_{\perp}) \gamma^5}
 | \mathbf{r}_{\perp}^{\prime}\rangle
 \end{eqnarray}
allows us to relate it to the propagator in the special case in Eq.~(\ref{G-b-parallel-B}); i.e.,
\begin{equation}
G(\omega,k_z;\mathbf{r}_{\perp}, \mathbf{r}_{\perp}^{\prime})
=e^{i (\mathbf{b}_{\perp}\cdot\mathbf{r}_{\perp}) \gamma^5}
G_{\parallel}(\omega,k_{z};\mathbf{r}_{\perp}, \mathbf{r}_{\perp}^{\prime})
e^{i (\mathbf{b}_{\perp}\cdot\mathbf{r}_{\perp}^{\prime}) \gamma^5}.
\end{equation}
By making use of this relation and taking into account that the result in Eq.~(\ref{GDn-new}) consists of
two separate chiral contributions,
$\bar{G}_{\parallel}\equiv \sum_{\chi=\pm}\bar{G}^{(\chi)}_{\parallel}{\cal P}_{\chi}$,
we derive the following explicit form for the propagator:
\begin{eqnarray}
G(\omega,k_z;\mathbf{r}_{\perp}, \mathbf{r}_{\perp}^{\prime})
&=&e^{i (\mathbf{b}_{\perp}\cdot\mathbf{r}_{\perp}) \gamma^5}
e^{i\Phi(\mathbf{r}_{\perp}, \mathbf{r}_{\perp}^{\prime})}
\int \frac{d^2\mathbf{k}_{\perp}}{(2\pi)^2}
e^{i \mathbf{k}_{\perp} \cdot (\mathbf{r}_{\perp}-\mathbf{r}_{\perp}^{\prime})}
\bar{G}_{\parallel}(\omega,k_{z},\mathbf{k}_{\perp})
e^{i (\mathbf{b}_{\perp}\cdot\mathbf{r}_{\perp}^{\prime}) \gamma^5} \nonumber \\
&=&
e^{i\Phi(\mathbf{r}_{\perp}, \mathbf{r}_{\perp}^{\prime})}
 \int \frac{d^2\mathbf{k}_{\perp}}{(2\pi)^2}
e^{i \mathbf{k}_{\perp}\cdot (\mathbf{r}_{\perp}-\mathbf{r}_{\perp}^{\prime})}
 \sum_{\chi=\pm}
\bar{G}^{(\chi)}_{\parallel}(\omega,k_{z},\mathbf{k}_{\perp}+\chi \mathbf{b}_{\perp})
{\cal P}_{\chi}.
\end{eqnarray}

\section{Integrals used in the derivation of the gap equation}
\label{AppIntegrals}

In the derivation of the gap equation, one encounters the following type of the energy integration:
\begin{equation}
\int_{-\infty}^{\infty} \frac{d\omega (\omega+\mu+A)}{\left[\omega+\mu +i\varepsilon \sign(\omega)\right]^2-B^2} =
i\pi \sign(\mu)\theta(\mu^2-B^2)- i\pi \frac{A}{B}\theta(B^2-\mu^2),
\label{Int-Delta-LLL-exp-01-A}
\end{equation}
where $A$ and $B$ are constants.

Also, one has to calculate the following two types of the momentum integrations:
\begin{eqnarray}
&&\int_{-\Lambda}^{\Lambda} dq_3f(q_3) \sign(\mu)\theta\left[\mu^2-v_F^2(b_z-q_3)^2-2nv_F^2|eB|\right]
=\sign(\mu)\theta\left[\mu^2-2nv_F^2|eB|\right] \int_{b_z-\Lambda^{\prime}}^{b_z+\Lambda^{\prime}} f(q_3)dq_3,\\
&&\int_{-\Lambda}^{\Lambda} dq_3f(q_3) \theta\left[v_F^2(b_z-q_3)^2+2nv_F^2|eB|-\mu^2\right]
=\theta\left[2nv_F^2|eB|-\mu^2\right] \int_{-\Lambda}^{\Lambda} dq_3f(q_3)
\nonumber \\
&&\hspace{2.75in}+\theta\left[\mu^2-2nv_F^2|eB|\right] \left(\int_{-\Lambda}^{b_z-\Lambda^{\prime}} dq_3f(q_3) +\int_{b_z+\Lambda^{\prime}}^{\Lambda} dq_3f(q_3)\right),
\end{eqnarray}
where $\Lambda^\prime \equiv \sqrt{(\mu/v_F)^2-2n|eB|}$.
In the two cases encountered in the gap equation, in particular, these lead to the following
results:
\begin{equation}
\int_{-\infty}^{\infty} d\omega \int_{-\Lambda}^{\Lambda} dq_3 \frac{\omega+\mu \pm s_{\perp}v_F(b_z-q_3)}
{(\omega+\mu)^2-v_F^2(q_3-b_z)^2-2nv_F^2|eB|}
=2i \pi \left[ \sign(\mu)\theta\left(\mu^2-2nv_F^2|eB|\right) \Lambda^{\prime}
\mp s_{\perp} b_z \right],
\end{equation}
and
\begin{eqnarray}
\int_{-\infty}^{\infty} d\omega \int_{-\Lambda}^{\Lambda} dq_3 \frac{-4v_F(b_y+q_y)}{ (\omega+\mu)^2-v_F^2(q_3-b_z)^2-2nv_F^2|eB|}
&=& 4i\pi(b_y+q_y)\Bigg[\theta\left[2nv_F^2|eB|-\mu^2\right] \ln\frac{2\Lambda^2}{n|eB|}
\nonumber\\
&+&\theta\left[\mu^2-2nv_F^2|eB|\right] \ln\frac{4 v_F^2 \Lambda^2 }{\left(|\mu|+ \sqrt{\mu^2-2nv_F^2|eB|}
\right)^2}
\Bigg].
\end{eqnarray}

\end{document}